\documentclass[10pt]{article}

\usepackage{amsmath}
\usepackage{amssymb}

\renewcommand{\thesection}{\arabic{section}}

\newtheorem{theorem}{Theorem}[section]
\newtheorem{prop}[theorem]{Proposition}
\newtheorem{proposition}[theorem]{Proposition}
\newtheorem{lemma}[theorem]{Lemma}

\newcommand{\qed}{\hbox{\hskip 1pt \vrule width 7pt height 7pt
            depth 1.5pt \hskip 1pt}}

\newcommand{\R}{{\BR}}
\newcommand{\D}{{\cal D}}
\newcommand{\BC}{{\mathbb C}}

\newcommand{\BR}{{\mathbb R}}
\newcommand{\bfone}{{\bf 1}}
\newcommand{\id}{{\bfone}}
\newcommand{\dom}{{\cal D}}
\newcommand{\cA}{{\cal{A}}}
\newcommand{\cD}{{\cal{D}}}
\newcommand{\cH}{{\cal{H}}}
\newcommand{\cHo}{{\cal{H}}_0}

\newcommand{\rIm}{{\rm{Im\, }}}

\newcommand{\rRe}{{\rm{Re\, }}}
\newcommand{\dist}{{\rm{dist\, }}}

\newcommand{\oS}{{\overline{S}}}
\newcommand{\oP}{{\overline{P}}}

\newcommand{\bs}{{\backslash}}

\newcommand{\thetabar}{{\overline{\theta}}}

\newcommand{\innerprod}[2]{\left\langle {#1}, {#2}\right\rangle}

\newcommand{\set}[2]{\left\{ #1\vphantom{#2} \right. \left| \vphantom{#1} #2 \right\}}
\newcommand{\norm}[1]{\left\| #1 \right\|}

\newcommand{\GL}{{\rm GL}}

\renewcommand{\d}{{\rm d}}
\newcommand{\bbbone}{\mathchoice {\rm 1\mskip-4mu l} {\rm 1\mskip-4mu l}
{\rm 1\mskip-4.5mu l} {\rm 1\mskip-5mu l}}
\newcommand{\lless}{<\!\!<}
\newcommand{\scalprod}[2]{\left\langle {#1}, {#2}\right\rangle}
\newcommand{\fer}[1]{(\ref{#1})}
\newcommand{\av}[1]{\left\langle{#1}\right\rangle}

\newcommand{\cF}{{\cal{F}}}
\newcommand{\cB}{{\cal{B}}}

\newcommand{\betamax}{\beta}

\renewcommand{\theequation}{\thesection.\arabic{equation}}
\let\subs\subsection
\renewcommand\subsection{\setcounter{equation}{0}
\gdef\theequation{\thesubsection.\arabic{equation}}\subs}
\let\sect\section
\renewcommand\section{\setcounter{equation}{0}
\gdef\theequation{\thesection.\arabic{equation}}\sect}

\begin{document}
\title{
Instability of Equilibrium States for Coupled Heat Reservoirs at Different Temperatures}

\author{M. Merkli\footnote{Department of Mathematics and Statistics, McGill University, 805 Sherbrooke W., Montreal, QC, H3A 2K6, Canada
(merkli@math.mcgill.ca)}
\thanks{Partly supported by an NSERC PDF and by the Institute of Theoretical Physics of ETH Z\"urich, Switzerland.}
\and M. M\"uck \footnote{Fachbereich Mathematik, Johannes
Gutenberg-Universit\"at, D-55128 Mainz, Germany
(mueck@mathematik.uni-mainz.de)}
\thanks{A part of the Ph.D. requirements, supported by DAAD under
grant HSP III.} \and I.M. Sigal \footnote{University of Notre Dame and University of Toronto}
\thanks{Supported by NSF under grant DMS-0400-526.}}
\date{July 8, 2005}
\maketitle

\begin{abstract}
We consider quantum systems consisting of a ``small'' system coupled to two reservoirs (called open systems). We show that such a system has no equilibrium states normal with respect to any state of the decoupled system in which the reservoirs are at different temperatures, provided that either the temperatures or the temperature difference divided by the product of the temperatures are not too small. 

Our proof involves an elaborate spectral analysis of a general class of generators of the dynamics of open quantum systems, including quantum Liouville operators (``positive temperature Hamiltonians'') which generate the dynamics of the systems under consideration. 
\end{abstract}

\setlength{\baselineskip}{15pt}
\section{Introduction}\label{intro1}

It seems obvious that a quantum system consisting of a small subsystem coupled to several reservoirs at different temperatures does not have an equilibrium state. However, such a result (a precise formulation of which we present in Section \ref{Sect3}) was proven only recently, in \cite{JP:NESS} for (two) fermionic heat baths at temperatures $T_1$ and $T_2$, under the condition
$
0<|g|<C \min\big(T_1,T_2, g_1(\Delta T)\big),
$
where $g$ is the interaction strength (coupling constant), $\Delta T =|T_1-T_2|>0$, and in \cite{DJ:RtEPauliFierz} for bosonic reservoirs, under the condition
$
0<|g|< g_2(\Delta T),
$ 
{\it uniformly} in $T_1,T_2\rightarrow 0$. Here $g_{1,2}(\Delta T)$ are some (implicit) functions which vanish in the limit $\Delta T\rightarrow 0$.
One of our goals is to prove absence of equilibria for small coupling constants, uniformly in $T_j\rightarrow 0$, {\it and uniformly in $\Delta T\downarrow 0$}. In this paper we take the first step in this direction by proving non-existence of equilibria under either of the following conditions 
\begin{itemize}
\item[--] $0< |g|< c[\min(T_1, T_2)]^{\frac{1}{2+\alpha}}$ (except possibly for a finite set of points) and  $|T_1^{-1}-T_2^{-1}|<c'$ for some $c'>0$,
\end{itemize}
or 
\begin{itemize}
\item[--] $0<|g| < c\left[\frac{|\Delta T|^2}{T_1T_2+|\Delta T|^2}\right]^{1/\alpha}$, 
\end{itemize}
where $\alpha=\frac{\mu-1/2}{\mu+1/2}$. Here, $c$ is an absolute constant and $\mu>1/2$ is a parameter describing the infra-red behaviour of interactions (see Condition (A) and Remark 2 in Section \ref{Sect3} below, and the next paragraph). In Section \ref{Sect_13} we sketch the strategy how to prove the instability of equilibrium states without temperature-dependent restrictions on the coupling strength. The detailed analysis of this is given in \cite{mms3}. 

Since the quantum excitations of the heat reservoirs (photons or phonons) are massless we have to deal with an infra-red technical problem. The severity of this problem is determined by the infra-red behaviour of coupling operators $G_j(k)$ entering the interaction term of the Hamiltonian, where $k\in {\mathbb R}^3$ is the momentum of photons (or phonons). 
Our results hold for $G_j(k)$ proportional, at $|k|\rightarrow 0$, to $|k|^p$, where $p$ can take the values $n+1/2$, with $n=0,1,2,\ldots$ ($p>\mu-1$, where $\mu$ is the parameter in the preceding paragraph). This is the same infra-red condition as in \cite{JP:NESS}, and it presents an improvement of the one in \cite{DJ:RtEPauliFierz}, since \cite{DJ:RtEPauliFierz} requires $p>2$, though with less restrictions on the regularity of $k\mapsto G_j(k)$.

Our approach is based on the characterization of equilibrium states in terms of eigenvectors corresponding to the eigenvalue zero of certain selfadjoint operators $L$, called {\it Liouville operators}, which act on the GNS representation Hilbert space (positive temperature Hilbert space) (see \cite{JP:QFII,BFS:RtE,JP:MathTheoryNEStQM,FM:TI}).

Parts of our techniques can be viewed as a perturbation theory in the temperatures, around $\delta\beta:=|T_1^{-1}-T_2^{-2}|=0$. This is a {\it singular perturbation theory} in the sense that the Hilbert spaces representations of the system for $\delta\beta=0$ and $\delta\beta>0$ are not normal with respect to each other (\cite{takesaki, BR1, BR2}).

Our techniques are applicable to a wide class of non-selfadjoint operators $K$, containing in particular the Liouville operators mentioned above, but also containing non-selfadjoint generators of the dynamics used in the examination of non-equilibrium stationary states (\cite{JP:NESS,mms2}). We thus carry out our analysis for this more general class of operators. 

In order to study the spectrum of the operators $K$, we develop a new type of spectral deformation, $K\mapsto K_\theta$, with a spectral deformation parameter $\theta\in {\mathbb C}^2$, which combines the deformations introduced in \cite{JP:QFII} and in \cite{BFS:RtE},
hence $\theta$ is in ${\mathbb C}^2$ rather than in ${\mathbb C}$.
(Such a combination was already mentioned in \cite{BFS:RtE}.) In
order to establish the desired spectral characteristics of the
operator family $K_\theta$, we use the method of the Feshbach map,
and perform the basic step of the spectral renormalization group approach as developed in
\cite{BCFS:SmoothFeshbach,BFS:QED,BFS:RG}.

Already a single application of the Feshbach map, considered in
this paper, yields the results mentioned above. Adapting ideas of \cite{BCFS:SmoothFeshbach,
BFS:RG, BFS:RtE} on the full renormalization group approach, the restriction on the temperatures can be removed.  We present in \cite{mms3} a detailed analysis of the RG to the specific model at hand. It relies on \cite{BFS:QED, BFS:RG,
BFS:RtE} and features some simplifications due to the specificity of our
problem and some recent developments \cite{BCFS:SmoothFeshbach}.

In contrast to the case of quantum Hamiltonians for zero temperature systems, the spectral theory of time-translation generators of open quantum systems is at an early stage of its development. Our paper is a contribution to this theory.

This paper is organized as follows. In Section 2 we describe our model and define the dynamics of it. (The definition of the dynamics is a somewhat subtle matter.) In Section 3 we give a precise
formulation of our assumptions, state the results and discuss
assumptions and results. In Section 4 we present the Araki-Woods construction which we use throughout this paper. In Section 5 we define a spectral deformation of a family of operators $K$ which contains the generator of the evolution, and we establish some basic analyticity and spectral properties of those operators.  In Section 6 we carry out a more refined spectral analysis, preparing for a proof of absence of normal invariant states, which is given in Section 7. Finally, in
Appendices \ref{appA}--\ref{appC'} we collect some technical results.

\section{Model and Mathematical Framework}
\label{Sect2}

We consider a system consisting of a particle system, described by
a Hamiltonian $H_p$ on a Hilbert space $\cH_p$, and two (thermal)
reservoirs, at inverse temperatures $\beta_1$ and $\beta_2$,
described by the Hamiltonians $H_{r1}$ and $H_{r2}$ acting on
Hilbert spaces $\cH_{r1}$ and $\cH_{r2}$, respectively. The full
Hamiltonian is
\begin{equation}\label{eq2.1}
H := H_0 + gv \ ,
\end{equation}
acting on the tensor product space $\cHo := \cH_p \otimes \cH_{r1}
\otimes \cH_{r2}$. Here
\begin{equation}
\label{hnot} 
H_0 := H_p \otimes \bfone \otimes \bfone + \bfone
\otimes H_{r1} \otimes \bfone + \bfone \otimes \bfone \otimes
H_{r2}
\end{equation}
is the unperturbed Hamiltonian, $v$ is an operator on $\cHo$
describing the interaction and $g\in{\mathbb R}$ is a coupling
constant.

For the moment we just require that $H_p$ is a self-adjoint
operator on $\cH_p$, with the property that ${\rm Tr\,}e^{-\beta H_p}<\infty$ (any $\beta>0$). The operators $H_{rj}$ describe free scalar
(or vector, if wished) quantum fields on $\cH_{rj}$, the bosonic Fock spaces over the one-particle space $L^2({\mathbb R}^3,d^3k)$, 
\begin{equation}
  H_{rj} = \int \omega(k) a^*_j(k) a_j(k) \, d^3k,
\label{fieldhamilt}
\end{equation}
where $a^*_j(k)$ and $a_j(k)$ are creation and annihilation
operators on $\cH_{rj}$ and $\omega(k) = |k|$ is the dispersion
relation for relativistic massless bosons. The interaction
operator is given by
\begin{equation}
\label{intop} 
v=\sum_{j=1}^2 v_j \mbox{\ \ \ with \ \ $v_j =
a_j(G_j) + a^*_j(G_j)$.}
\end{equation}
Its choice is motivated by standard models of particles
interacting with the quantized electromagnetic field or with
phonons.

Here,  $G_j : k \mapsto G_j(k)$ is a map from $\R^3$ into ${\cal
B}({\cal H}_p)$,  the algebra of bounded operators on $\cH_p$, and
\begin{equation} \label{ahah}
 a_j(G_j) := \int  G_j(k)^* \otimes a_j(k) \, d^3k \qquad
 \text{and} \qquad a_j^*(G_j) := a_j(G_j)^*.
\end{equation}

If the coupling operators $G_j$  are such that
\begin{equation} \label{eqn_cond_sa}
g^2\int\limits_{\R^3} \left( 1 + |k|^{-1} \right) \norm{G_j(k)}^2
\, dk \qquad \text{is sufficiently small},
\end{equation}
then the operator $H$ is self-adjoint (see e.g. \cite{BFS:RtE}).

Now we set up a mathematical framework for non-equilibrium
statistical mechanics. Operators on the Hilbert space ${\cal H}_0$
will be called observables. (Strictly speaking only certain
self-adjoint operators on ${\cal H}_0$ are physical observables.)
As an algebra of observables describing the system we take the
$C^*$-algebra
\begin{equation} 
\label{calA}
{\cal A}={\cal B}({\cal H}_p)\otimes{\frak W}(L^{2}_0)\otimes{\frak W}(L^{2}_0),
\end{equation}
where ${\frak W}(L^{2}_0)$ denotes the Weyl CCR algebra over the
space $L^{2}_0 := L^2({\mathbb R}^3, (1+|k|^{-1}) d^3k)$. 
States of the system are positive linear functionals, $\psi$, on the algebra ${\cal A}$ normalized
as $\psi(\bfone)= 1$.

The reason we chose ${\cal A}$ rather than ${\cal B}({\cal H}_0)$
is that the algebra ${\cal A}$ supports states in which each
reservoir is at a thermal equilibrium at its own temperature. More
precisely, consider the evolution for the $i$-th reservoir given
by
\begin{equation}
 \alpha_{ri}^t (A) := e^{iH_{ri}t} A e^{-iH_{ri}t}  .
\end{equation}
Then there are stationary states on the $i$-th reservoir algebra of
observables, $\frak{W}(L^{2}_0)$, which describe thermal
equilibria.
These states are parametrized by the inverse temperature $\beta$
and their
generating functional is given by
\begin{equation}
\label{mm31}
\omega^{(\beta)}_{ri} \left(W_{i}(f)\right)=
\exp\left\{-\frac{1}{4}\int_{{\mathbb R}^3}\frac{e^{\beta
|k|}+1}{e^{\beta
|k|}-1}|f(k)|^2 d^3k\right\}, 
\end{equation}
where  $W_j(f):=e^{i\phi_j(f)}$, with
$\phi_j(f):=\frac{1}{\sqrt 2}\left(a^*_j(f)+a_j(f)\right)$, is a Weyl operator, 
see e.g.
\cite{BR2}. The choice of the space $L^{2}_0$ above is dictated by
the need to have the r.h.s. of this functional finite. These
states are characterized by the KMS condition and are called
$(\alpha^t_{ri} , \beta )$-KMS states.

{\it Remark.\ } It is convenient to define states $\psi$ on products
$a^{\#}(f_1) \ldots a^{\#}(f_n)$ of the creation and annihilation
operators, where $a^\#$ is either $a$ or $a^*$.  This is done
using $s$-derivatives of its values on the Weyl operators
$W(s_1f_1) \ldots W(s_nf_n)$ (see \cite{BR2}, Section 5.2.3 and (\ref{eq2.15})).

Consider states (on ${\cal A}$) of the form 
\begin{equation}\label{eq2.2}
\omega_0 := \omega_p \otimes\omega_{r1}^{(\beta_1)} \otimes\omega_{r2}^{(\beta_2)},
\end{equation}
where $\omega_p$ is a state of the particle system and $\omega^{(\beta)}_{ri}$ is the $(\alpha^t_{ri} , \beta
)$-KMS state of the $i$-th reservoir. The set of states which are normal w.r.t. $\omega_0$ is the same for any choice of $\omega_p$. A state $\psi$ which is normal w.r.t. $\omega_0$ (i.e., which is represented by a density matrix $\rho$ in the GNS representation $({\cal H}, \pi,\Omega_0)$ of $({\frak A},\omega_0)$, according to $\psi(A)={\rm Tr}(\rho\pi(A))$) will be called a {\it $\beta_1\beta_2$-normal state}. 

In the particular case $\omega_p(\cdot)={\rm Tr}(e^{-\beta_p H_p}\,\cdot)/{\rm Tr}(e^{-\beta_p H_p})$ we call $\omega_0$ a {\it reference state}.

The Hamiltonian $H$ generates the dynamics of observables $A\in
{\cal B}({\cal H}_0)$ according to the rule
\begin{equation}\label{eq2.6}
A \mapsto \alpha^t (A) := e^{iHt} A e^{-iHt} \ .
\end{equation}
Eqn (\ref{eq2.6}) defines a group of *-automorphisms of
$\cB(\cH_0)$. However, $\alpha^t$ is not expected to map the
algebra ${\cal A}$ into itself. To circumvent this problem we define the
interacting evolution of states on ${\cal A}$ by using the Araki-Dyson 
expansion. Namely, for a state $\psi$ on the
algebra $\cA$ normal w.r.t. the state $\omega_0$, we define the
evolution by
\begin{equation} 
\label{dyn1}
\psi^t(A):=\lim_{n \rightarrow \infty} \sum_{m=0}^\infty(ig)^m\int_0^tdt_1
\cdots\int_0^{t_{m-1}} dt_m \ 
\psi_n^{t,t_1,\ldots,t_m}(A),
\end{equation}
where the term with $m=0$ is $\psi(\alpha_0^{t}(A))$, and, for $m\geq 1$,
\begin{equation*}
\psi_n^{t,t_1,\ldots,t_m}(A) :=
\psi\left([\alpha_0^{t_m}(v_{n}),[
\cdots [\alpha_0^{t_1}(v_{n}),\alpha_0^{t}(A)]\cdots]]\right).
\end{equation*}
Here, $v_n \in
\cA$ is an approximating sequence for the operator $v$,
satisfying the relation
\begin{equation}
\label{eq2.13}
 \lim_{n\rightarrow\infty}\omega_{0}(A^{*}(v_{n}^* - v^*)(v_n - v)A)=  0, 
\end{equation}
$\forall A \in \cA$ of the form $A=B\otimes W_1(f_1)\otimes W_2(f_2)$ with $B\in{\cal B}({\cal H}_0)$, $f_{1,2} \in
L^{2}_0$. Such a sequence is constructed as follows. Let $\{e_m\}$  be an 
orthonormal basis of $L^{2}_0$.
We define the approximate creation operators
\begin{equation} \label{eq2.14}
 a_{j,n}^*(G_j) = \sum_{m=1}^{M}\langle{e}_{m},G_{j}\rangle
 b_{j,\lambda}^*(e_{m}),
 \end{equation}
where $n = (\lambda,M)$, 
and, for any $f\in L^{2}({\mathbb R}^3)$ and $\lambda > 0$,
\begin{equation}
\label{eq2.15}
b^*_{j,\lambda}(f):=\frac{\lambda}{\sqrt 2 i}\left\{
W_j(f/\lambda) - \bfone - i W_j(if/\lambda) + i \bfone \right\}.
\end{equation}
Similarly we define the approximate annihilation operators
$a_{j,n}(G_j)$. Via the above construction we obtain the family
of interactions $v_n\in {\cal A}$. Using \fer{mm31}, one easily shows that \fer{eq2.13} is satisfied. 


In Appendix \ref{appA} we show that under condition \fer{eq2.13} 
the integrands on the r.h.s. of \fer{dyn1} are continuous
functions in $t_1,\ldots,t_m$, that the series is absolutely convergent and that the limit exists and is independent of the approximating sequence $v_n$.

A $\beta_1 \beta_2$-normal state $\psi$ is called invariant (under the interacting dynamics), or stationary, if $\psi^t(A)=\psi(A)$ for all $A\in\cal A$, $t\in{\mathbb R}$, see \fer{dyn1}. Our goal is to show that, if $\beta_1\neq \beta_2$, then there are no $\beta_1 \beta_2$-normal states which are invariant. In particular, there are no equilibrium states (see Theorem \ref{thm2.1}).

To pass to a Hilbert space framework one uses the 
GNS representation of $({\cal A},\omega_0)$, where 
$\omega_0$ is given in  (\ref{eq2.2}):
\begin{equation*}
(\cA , \omega_0) \to (\cH , \pi , \Omega_0).
\end{equation*}
Here $\cH$, $\pi$ and $\Omega_0$ are a Hilbert space, a
representation of the algebra $\cA$ by bounded operators on $\cH$,
and a cyclic element in $\cH$ (i.e. $ \overline{\pi(\cA) \Omega_0}
= \cH)$ s.t.
\begin{equation*}
\omega_0(A) = \innerprod{ \Omega_0 }{ \pi (A) \Omega_0} \ .
\end{equation*}
(In this paper we use the Araki-Woods GNS representation with $\omega_p (A) :=
{\rm Tr}(e^{-\beta_pH_p}A) / {\rm Tr} (e^{-\beta_pH_p})$ in \fer{eq2.2}, see Section \ref{sect:araki-woods}.)

With the free evolution $ \alpha_0^{t}(A):=e^{itH_0}Ae^{-itH_0}$
one associates the unitary one-parameter group, $U_0 (t) = e^{i t L_{0}}$, on
$\cH$ s.t.
\begin{equation}
\pi (\alpha^t_0 (A) ) = U_0(t) \pi (A) U_0 (t)^{-1}
\label{2.25a}
\end{equation}
and $U_0 (t) \Omega_0 = \Omega_0 $. Define the {\it standard Liouville operator}
\begin{equation}
L:=L_0+g\pi(v)-g\pi'(v),
\label{standardliouvillian}
\end{equation}
defined on the dense domain ${\cal D}(L_0)\cap{\cal D}(\pi(v))\cap{\cal D}(\pi'(v))$. Here, $\pi(v)$ and $\pi'(v)$ can be defined either using 
explicit formulae for $\pi$ and $\pi'$ in the Araki-Woods representation given below, or 
by using the approximation $v_n\ \in \mathcal{A}$ of $v$, constructed above. 
By the Glimm-Jaffe-Nelson commutator
theorem, the operator $L$ is essentially self-adjoint; we denote its
self-adjoint closure again by $L$. The operator
$L$ generates the one-parameter group of $*$automorphisms $\sigma^t$ on the von Neumann algebra
$\pi({\cal A})''$ (the weak closure of  $\pi({\cal A})$), 
\begin{equation} 
\label{eq2.26}
\sigma^{t}(B):=e^{itL} B e^{-itL},
\end{equation}
where $B \in \pi(\cA)''$. Let $\psi$ be a state on the algebra
$\cA$ normal w.r.t. the state $\omega_0$, i.e.
\begin{equation}\label{eq2.27}
\psi(A)= {\rm Tr} (\rho  \pi(A))
\end{equation}  
for some positive trace class operator $\rho$ on $\cH$ of trace one. It is shown
in Appendix \ref{appA} that for $\psi$ as above the limit on the r.h.s. of 
\fer{dyn1} exists and equals 
\begin{equation}\label{eq2.28}
\psi^t(A)= {\rm Tr} (\rho \sigma^{t}(\pi(A))).
\end{equation}
In particular, the limit is independent of the choice of the approximating family $v_n$. 

The following result connects the existence of normal invariant states to spectral properties of the standard Liouvillian $L$:
\begin{theorem}[\cite{JP:MathTheoryNEStQM, FM:TI}]
\label{ninvspecthm}
A normal $\sigma^t$-invariant state on $\pi({\cal A})''$ exists if and only if zero is an eigenvalue of $L$.
\end{theorem}

In order to obtain rather subtle spectral information on the
operator $L$, we develop a new type of spectral deformation,
$L\mapsto L_\theta$, with a spectral deformation parameter
$\theta\in {\mathbb C}^2$. This deformation has the property that zero is an eigenvalue of $L$ if and only if zero is an eigenvalue of $L_\theta$, for $\theta\in ({\mathbb C}_+)^2$. We then investigate the spectrum of $L_\theta$, using a Feshbach map iteratively.

\section{Assumptions and Results}
\label{Sect3}

For our analysis we need conditions considerably stronger than
(\ref{eqn_cond_sa}). In order to formulate them, we
first introduce some definitions. We refer the reader to the
remarks at the end of this section for a discussion of the definitions and conditions. 

We define the map $\gamma: L^2({\mathbb R}^3) \to
L^2({\mathbb R}\times S^2)$,
\begin{equation}\label{eq_2_8}
  (\gamma f)(u,\sigma) = 
  \sqrt{|u|}\
\left\{
\begin{array}{ll}
 f(u\sigma), & u\geq 0,\\
 -\overline{f}(-u\sigma), & u<0.
\end{array}
\right.
\end{equation}
Let $j_\theta(u) = e^{\delta {\rm sgn}(u)} u + \tau$ for $\theta =
(\delta, \tau) \in \BC^2$ and $u \in \BR$ (see \fer{eqn_A.25}) and
define $({\gamma_{\theta}} f)(u, \sigma) = (\gamma
f)(j_\theta(u), \sigma)$, for $f\in L^2(\BR  \times S^2)$, $\theta\in{\mathbb R}^2$.

We extend the maps  $\gamma$ and ${\gamma_{\theta}}$
to operator valued functions in the obvious way. Now, we are ready
to formulate our assumptions.

\begin{itemize}
\item[(A)] {\it  Analyticity.\ } 
{}For $j=1,2$ and every fixed $(u,\sigma)\in{\mathbb R}\times S^2$, the maps 
\begin{equation}
\theta \mapsto (\gamma_\theta G_j)(u,\sigma)
\end{equation}
from ${\mathbb R}^2$ to the bounded operators on ${\cal H}_p$ have analytic continuations to
\begin{equation}\label{eq_2_11.1}
\Big\{ (\delta, \tau) \in \BC^2 \big| |\rIm\delta| < \delta_0,
|\tau|
           < \tau_0 \Big\} \ ,
\end{equation}
for some $\delta_0$, $\tau_0 > 0$, $\frac{\tau_0}{\cos\delta_0} \le \frac{2\pi}{\beta}$, where 
$\beta=\max(\beta_1,\beta_2)$. Moreover, 
\begin{equation}
\label{eqn_cond_B}
\norm{G_j}_{\mu, \theta} := \sum_{\nu=1/2,\mu}
\left[\ \int\limits_{\BR \times S^2}
\left\|\gamma_\theta\left[ \frac{\sqrt{|u|+1}}{|u|^\nu}  G_j\right](u,\sigma)\right\|^2 du d\sigma\right]^{1/2}<\infty,
\end{equation}
for some fixed $\mu>1/2$.


  \item[(B)] {\it  Fermi Golden Rule Condition.}
     \begin{eqnarray} 
\label{eqn_2.7}
       \gamma_{0j} := \min_{0 \le n < m \le N-1} \int_{{\mathbb R}^3} \delta(|k|-|E_{nm}|)\  |G_j(k)_{nm}|^2 d^3k > 0,\ \ \ j=1,2,
     \end{eqnarray}
 where $G_j(k)_{mn} :=
 \innerprod{\varphi_m}{G_j(k)\varphi_n}$,  $\varphi_n$ are
 normalized eigenvectors of $H_p$ corresponding to the eigenvalues
 $E_n$, $n=0,\ldots,N-1$, and $\delta$ is the Dirac delta distribution.
\end{itemize}
For some of our results, we impose the additional condition
\begin{itemize}
\item[(C)] {\it Simplicity of spectrum of $H_p$.\ } The eigenvalues of the particle Hamiltonian $H_p$ are simple. 
\end{itemize}

Let
\begin{equation}
\sigma :=\min\left\{|\lambda-\mu|\ |\ \lambda,\mu\in\sigma(H_p), \lambda\neq\mu\right\}.
\label{gap}
\end{equation}
Define 
\begin{equation}
g_0:= C\sigma^{1/2} \sin(\delta_0)\left[(1+\beta_1^{-1/2}+\beta_2^{-1/2}) \max_j \sup_{|\theta|\leq\theta_0}\|G_j\|_{1/2,\theta}\right]^{-1},
\label{g_0}
\end{equation}
where $C$ is a constant depending only on $\tan\delta_0$, and set
\begin{equation}
g_1:=\min\left((g_0)^{1/\alpha}, [\min(T_1,T_2)]^{\frac{1}{2+\alpha}}\right).
\label{g_1}
\end{equation}

{\it Remarks.\ }
1) The map \fer{eq_2_8} has the following
origin. In the positive-temperature representation of the CCR (the
Araki-Woods representation on a suitable Hilbert space, see
Appendix A), the interaction term $v_j$ is
represented by $a_j(\widetilde\gamma_{\beta_j}G_j)+ a^*_j(\widetilde\gamma_{\beta_j}G_j)$, where 
\begin{equation}
\widetilde\gamma_\beta := \sqrt{\frac{u}{1-e^{-\beta u}}}\ \gamma.
\label{gammatilde}
\end{equation}

2)\ A class of interactions satisfying
Condition~(A) is given by $G_j(k)=g(|k|) G$, where
$g(u)=u^p e^{-u^2}$, with $u\geq 0$, $p=n+1/2$,
$n=0,1,2,\ldots$, and $G=G^*\in{\cal B}({\cal H}_p)$. A
straightforward estimate gives that the norms (\ref{eqn_cond_B}) have the bound
\begin{equation}
 \norm{G_j}_{\mu, \theta} \leq C ||G||,\label{b1}
\end{equation}
provided
$\mu < p + 1$, where the constant $C$ does not depend on the
inverse temperatures, nor on $\theta$ varying in any compact set.

The restriction $p=n+1/2$ with
$n=0,1,2,\ldots$ comes from the requirement of translation
analyticity (the $\tau$--component of $\theta$), which appears
also  in \cite{JP:NESS}.

3)\ 
The condition $\tau_0/\cos\delta_0<2\pi /\beta$ after \fer{eq_2_11.1} guarantees that the square root in \fer{gammatilde} is analytic in translations $u\mapsto u+\tau$.

4)\ Condition (C) guarantees that the {\it level-shift operators} of the system have certain technical features which facilitate the analysis (see also Proposition \ref{absprop2} and \cite{BFS:RtE}). We believe that this condition can be removed.

Our result on instability of normal stationary states is

\begin{theorem}
\label{thm2.1}
Assume conditions (A), (B) and (C) are obeyed for some
$0<\beta_1,\beta_2<\infty$, $\mu>1/2$, and set $\alpha=(\mu-1/2)/(\mu+1/2)$. Assume $\delta\beta:=|\beta_1-\beta_2|\neq 0$. There are constants $c,c',c''$ s.t. if either of the two following conditions hold,

1. $0<|g|<cg_1$, $\delta\beta<c'$, $\|G_1-G_2\|<c'$, and $g$ avoids possibly finitely many values in the set $\{0<|g|<cg_1\}$, or

2. $0<|g|<c'' \min\left((g_0)^{1/\alpha}, \big[\min_j(\gamma_{0j})\textstyle{\frac{|\delta\beta|^2}{1+|\delta\beta|^2}}\big]^{1/\alpha}\right)$,

\noindent
then there are no normal $\sigma^t$-invariant states on $\pi({\cal A})''$.
\end{theorem}

{\it Remarks.\ } 5) Using Araki's theory of perturbation of KMS states (c.f. \cite{DJP:PertTheoryWDynamics}) it is not hard to show that if the reservoir-temperatures are equal, then the system has an equilibrium state. 

6) By an analyticity argument one can show that the result 1. holds for all but a discrete set of values of $\delta\beta$ and  $\|G_1-G_2\|$.

7) We will remove the ``high temperature'' restriction $|g|<c[\min(T_1,T_2)]^{\frac{1}{2+\alpha}}$, \fer{g_1}, in \cite{mms3}; see the end of Section 7 for the relevant ideas.


\section{Araki-Woods representation and Liouville operators}
\label{sect:araki-woods}

In this section we present the explicit GNS representation 
provided by the Araki-Woods construction, which is used in  
our analysis 
(see \cite{BFS:RtE, JP:QFII, BR1, BR2} for details and \cite{AW,
HHW} for original papers).
In the Araki-Woods GNS representation the (positive temperature)
Hilbert space is given by
\begin{equation}\label{eqn_6_1}
  \cH = \cH^p \otimes \cH^r,
\end{equation}
where $\cH^p = \cH_p \otimes \cH_p$ and $\cH^r = \cH^{r1} \otimes
\cH^{r2}$ with
\begin{equation}\label{eqn_6_2}
  \cH^{rj} = {\cal H}_{rj} \otimes {\cal H}_{rj}.
\end{equation}
We denote by $a^{\#}_{\ell, j}(f)$ (resp., $a^{\#}_{r, j}(f)$) the
creation and annihilation operators which act on the left (resp.,
right) factor of (\ref{eqn_6_2}). They are related to the zero
temperature creation and annihilation operators $a_j^{\#}(f)$ by
\begin{equation}\label{eq_AWrepr}
  \pi(a_j(f)) = a_{\ell j} (\sqrt{1+\rho_j} \, f) +
  a^*_{r j} (\sqrt{\rho_j} \, \bar{f})
\end{equation}
and
\begin{equation}\label{eq_AWrepr'}
  \pi'(a_j(f)) = a^*_{\ell j} (\sqrt{\rho_j} \, f) +
  a_{r j} (\sqrt{1+\rho_j} \, \bar{f})
\end{equation}
where $\rho_j \equiv \rho_j(k) = (e^{\beta_j \omega(k)} - 1)^{-1}$
with $\omega(k) = |k|$. Finally, we denote $\Omega_r :=
\Omega_{r1} \otimes \Omega_{r2}$, where $\Omega_{rj} :=
\Omega_{rj,\ell} \otimes \Omega_{rj,r}$ are the vacua in
$\cH^{rj}$. Thus, $\Omega_r$ is the vacuum in $\cH^r$.

Definition (\ref{eq2.2}) and our choice of $\omega_p$ made at the
beginning of this section
imply that
\begin{equation}\label{eq_6_3}
  \Omega_0 = \Omega_p \otimes \Omega_r \quad \text{with} \quad
  \Omega_p \equiv \Omega^p_{\beta_p} = \frac{\sum_j e^{-\beta_pE_j/2}
  \varphi_j \otimes \varphi_j}{[\sum_j e^{-\beta_pE_j}]^{1/2}},
\end{equation}
where, recall, $E_j$ and $\varphi_j$ are the eigenvalues and
normalized eigenvectors of $H_p$.

The self-adjoint operator $L_0$ generating the free
evolution, $U_0(t)$, defined in \fer{2.25a}, is of the form $L_0 = L_p \otimes \id^r + \id^p
\otimes L_r$ with $L_r = \sum_{j=1}^2 L_{rj}$. The operator $L_p$
has the standard form
\begin{equation*}
  L_p = H_p \otimes \id_{p} - \id_{p} \otimes H_p
\end{equation*}
and the operators $L_{rj}$ are as follows
\begin{equation*}
  L_{rj} = \int \omega(k) \left( a^*_{\ell,j}(k) a_{\ell,j}(k)
     - a^*_{r,j}(k) a_{r,j}(k) \right) \,d^3k.
\end{equation*}
A standard argument shows that the spectrum of the operator $L_0$
fills the axis $\BR$ with the thresholds and eigenvalues located
at $\sigma(L_{p})$ and with $0$ an eigenvalue of multiplicity at least $\dim
H_p$ and at most $(\dim H_p)^2$ (depending on the degeneracy of the spectrum of $L_p$).

\section{A class of Liouville operators and their Spectral Deformation}
\label{Sect6}

To investigate the point spectrum of the self-adjoint Liouvillian $L$ we perform a complex deformation of the operator $L$, producing a family of operators $L_\theta$, $\theta\in {\mathbb C}^2$, with the property $L_{\theta=0}=L$ and s.t. $L_\theta$ is unitarily equivalent to $L$ for $\theta\in{\mathbb R}^2$. We investigate the spectrum of $L_\theta$ for complex $\theta$ which we relate to the properties of $L$ that are of interest to us. In this section we construct the family $L_\theta$ and establish some global spectral and analyticity properties. In the next section we give a finer description of the spectrum of $L_\theta$.

In fact, the analysis of both this section and the next one works for a general class of operators which are of the form
\begin{equation}
K:= L_0 +gI,\ \ \ I:=U-W',
\label{K}
\end{equation}
where $U=\pi(u)$ and $W'=\pi'(w)$, with operators $u,w$ of the form
\begin{eqnarray}
u &=& \sum_{j=1,2} \left\{a_j^*(G_{j1}) + a_j(G_{j2})\right\}\label{u}\\
w &=& \sum_{j=1,2} \left\{a_j^*(G_{j3}) + a_j(G_{j4})\right\}\label{w}.
\end{eqnarray}
If 
\begin{equation}
G_{jk}=G_j,\ \ \mbox{for $k=1,\ldots 4$ and $j=1,2$,}
\label{selfadjoint}
\end{equation}
then the operator $K$ reduces to the standard Liouville operator $L$, \fer{standardliouvillian}. We carry out the analysis for the more general class of operators $K$ since they are needed in the construction of non-equilibrium stationary states, \cite{mms2}. Note that in general, $K$ is not a normal operator.

For the spectral analysis of the operators $K$ we replace condition (A) by condition (AA) below, which reduces to (A) for self-adjoint $K$. For a scalar function $f(u,\sigma)$ and $k=1,3$, set
\begin{equation}
\gamma(fG_{jk})(u,\sigma):= |u|^{1/2}
\left\{
\begin{array}{ll}
f(u,\sigma) G_{jk}(u\sigma), & u\geq 0\\
-\overline{f}(-u,\sigma) G^*_{j(k+1)}(-u\sigma), & u<0
\end{array}
\right.
\label{gamma2}
\end{equation}
and define $\gamma_\theta(fG_{jk})$ as after \fer{eq_2_8} (if \fer{selfadjoint} holds then \fer{gamma2} coincides with $(\gamma G_j)(u,\sigma)$ as defined by \fer{eq_2_8}). 
\begin{itemize}
\item[(AA)] {\it Analyticity (non-selfadjoint case).\ } 
{}For $j=1,2$, $k=1,3$, and for every fixed $(u,\sigma)\in{\mathbb R}\times S^2$, the maps 
\begin{equation}
\theta \mapsto (\gamma_\theta G_{jk})(u,\sigma)
\end{equation}
from ${\mathbb R}^2$ to the bounded operators on ${\cal H}_p$ have analytic continuations to
\begin{equation}\label{eq_2_11}
\Big\{ (\delta, \tau) \in \BC^2 \big| |\rIm\delta| < \delta_0,
|\tau|
           < \tau_0 \Big\} \ ,
\end{equation}
for some $\delta_0$, $\tau_0 > 0$, $\frac{\tau_0}{\cos\delta_0} \le \frac{2\pi}{\beta}$, where 
$\beta=\max(\beta_1,\beta_2)$. Moreover, 
\begin{equation}
\label{eqn_cond_B2}
\norm{G_j}_{\mu, \theta} := \sum_{k=1,3} \sum_{\nu=1/2,\mu}
\left[\ \int\limits_{\BR \times S^2}
\left\|\gamma_\theta\left[ \frac{\sqrt{|u|+1}}{|u|^\nu}  G_{jk}\right](u,\sigma)\right\|^2 du d\sigma\right]^{1/2}<\infty,
\end{equation}
for some fixed $\mu>1/2$.
\end{itemize}

If \fer{selfadjoint} holds then condition (AA) coincides with condition (A). 
One shows that $K$ is a closed operator on the dense domain ${\cal D}(L)\cap{\cal D}(U)\cap{\cal D}(W')$. 

In order to carry
out the spectral analysis of the operator $K$, which we begin in
this section, we use the specifics of the Araki-Woods representation. They were not used in an essential way for the developments up to this section.

As a complex deformation we choose a combination of the complex
dilation used in \cite{BFS:RtE} and complex translation due to
\cite{JP:QFII} (see \cite{BFS:RtE}, Section V.2 for a sketch of
the relevant ideas).

First we define the group of dilations. Let $\hat{U}_{d, \delta}$
be the second quantization of the one-parameter group
\begin{equation*}
  u_{d, \delta} : f(k) \to e^{3\delta/2} f(e^{\delta}k)
\end{equation*}
of dilations on $L^2(\BR^n)$. This group acts on creation and
annihilation operators $a_r^{\#}(f)$ on the Fock space, $\cH_r$,
according to the rule
\begin{equation} \label{eqn_6_3}
  \hat{U}_{d, \delta} a_r^{\#}(f) \hat{U}_{d, \delta}^{-1} =
  a_r^{\#}(u_{d, \delta}f), \qquad \hat{U}_{d, \delta}
  \Omega_{rj} = \Omega_{rj}.
\end{equation}
We lift this group to the positive-temperature Hilbert space,
 (\ref{eqn_6_1}), according to the formula
\begin{equation} \label{eqn_6_4}
  U_{d, \delta} =  \id_{p} \otimes \id_{p} \otimes
  \hat{U}_{d, \delta} \otimes \hat{U}_{d, -\delta} \otimes
\hat{U}_{d, \delta} \otimes \hat{U}_{d, -\delta}.
\end{equation}
Note that we could dilate each reservoir by a different amount.
However, this does not give us any advantage, so to keep notation
 simple we use one dilation parameter for both
reservoirs. We record for future reference how the group $U_{d,
\delta}$ acts on the Liouville operator $L_0$ and the
positive-temperature photon number operator $N := \sum_{j=1}^2N_j$,
where
\begin{equation}
  N_j: = \int \left[ a^{\ast}_{\ell,j}(k) a_{\ell,j}(k) + a^{\ast}_{r,j}(k) a_{r,j}(k) \right] \,
  d^3k,
\end{equation}
and where the operators $a_{\{\ell,r\},j}^{\#}(k)$ were introduced after \fer{eqn_6_2}. 
We have (below we do not display the identity operators):
\begin{equation} \label{eqn6_6}
  U_{d, \delta} L_{rj} U_{d, \delta}^{-1} = \cosh(\delta) L_{rj}
  + \sinh(\delta) \Lambda_j,
\end{equation}
where $\Lambda_j$ is the positive operator on the $j$th reservoir
Hilbert space given by
\begin{equation}
  \Lambda_j = \int \omega(k) \left( a^{\ast}_{\ell,j}(k) a_{\ell,j}(k) + a^{\ast}_{r,j}(k) a_{r,j}(k) \right)
  \, d^3k,
\end{equation}
and
\begin{equation} \label{eqn6_8}
  U_{d, \delta} N_j U_{d, \delta}^{-1} = N_j.
\end{equation}

Now we define a one-parameter group of translations. It can be
defined as one-parameter group arising from transformations of the
underlying physical space similarly to the dilation group. This is
done in Appendix \ref{App_Repr}. Here we define the translation
group by means of the selfadjoint generator $T := \sum_{j=1}^2 T_j$, where
\begin{equation} \label{eqn_6_12}
  T_j = \int \left[ a^{\ast}_{\ell,j}(k) \vartheta a_{\ell,j}(k)
       + a^{\ast}_{r,j}(k)\vartheta a_{r,j}(k) \right] \,
       d^3k.
\end{equation}
Here, $\vartheta = i (\hat{k} \cdot \nabla + \nabla \cdot \hat{k})$
with $\hat{k} = k/|k|$. Since $\vartheta$ is a self-adjoint operator
on $L^2(\R^3)$, the operators $T_j, j=1,2$, and therefore the
operator $T$, are self-adjoint as well. We define the
one-parameter group of translations as
\begin{equation} \label{eqn_6_13}
  U_{t, \tau} := \bfone_p \otimes \bfone_p \otimes
  e^{i\tau T}.
\end{equation}
Eqns.~(\ref{eqn_6_12}) - (\ref{eqn_6_13}) imply the following
expressions for the action of this group on the Liouville
operators:
\begin{equation} \label{eqn6_11}
  U_{t, \tau} L_{rj} U_{t, \tau}^{-1} = L_{rj} + \tau N_j.
\end{equation}
Observe that neither the dilation nor the translation group
affects the particle vectors, and that $U_{t,\tau}N_j U_{t,\tau}^{-1}=N_j$. 

Now we want to apply the product of these transformations to the
full operator $K = L_0 + gI$, \fer{K}. Since the dilation and translation
transformations do not commute we have to choose the order in
which we apply them. The operator $\Lambda = \sum_{j}
\Lambda_j$ is not analytic under the translations, while the
operator $N$ is analytic under dilations. Thus we apply first the
translation and then the dilation transformation, and define the
combined translation-dilation transformation as
\begin{equation}
  U_{\theta} = U_{d, \delta} U_{t, \tau}
\end{equation}
where $\theta = (\delta, \tau)$. In what follows we will use the
notation $|\theta| = (|\delta|, |\tau|)$, $\rm{Im} \theta =
(\rm{Im}\delta, \rm{Im}\tau)$, and similarly for $\rm{Re}\theta$,
and
\begin{equation}
  \rm{Im} \theta > 0 \quad \iff \quad \rm{Im}\delta > 0 \wedge
  \rm{Im}\tau > 0.
\end{equation}

Now we are ready to define a complex deformation of the operator
$K$. On the set $D(\Lambda)\cap D(N)$ we define for $\theta \in
\BR$
\begin{equation}
  K_{\theta } := U_{\theta } K U_{\theta}^{-1}.
\end{equation}
Recalling the decomposition $K = L_0 + gI$, \fer{K}, where $L_0:=L_p+L_r$, $L_r :=
\sum_{j=1}^2 L_{rj}$ and $I = U - W'$, we have
\begin{equation} \label{eq6.18}
K_{\theta} = L_{0,\theta } + gI_{\theta},
\end{equation}
where the families $L_{0, \theta}$ and $I_{\theta}$ are defined
accordingly. Due to Eqns.~(\ref{eqn6_6}), (\ref{eqn6_8}) and
(\ref{eqn6_11}) we have:
\begin{equation} 
\label{eqn6_16}
L_{0, \theta} = L_p + \cosh(\delta) L_r +\sinh(\delta)\Lambda +
\tau N,
\end{equation}
where $\theta = (\delta, \tau)$, and $\Lambda = \sum_{j=1}^2
\Lambda_j$. An explicit expression for the family $I_\theta$ is
given in Appendix \ref{b2} (see Eqns \fer{itheta} and \fer{A2.7}).

Of course the operator families above are well defined for real
$\theta$. Our task is to define them as analytic families on the strips
\begin{equation} \label{eqn_6.17}
S_{\theta_0}^\pm = \left\{\theta \in \BC^2 | 0<\pm \rm{Im} \theta
< \theta_0 \right\}
\end{equation}
where $\theta_0 = (\delta_0, \tau_0) > 0$ is the same as in
Condition (AA). Recall that the inequality $\pm\rm{Im}\theta <
\theta_0$ is equivalent to the following inequalities:
$\pm\rm{Im}\delta < \delta_0$ and $\pm\rm{Im}\tau < \tau_0$. (The fact that analyticity in a neighbourhood of a fixed $\theta\in S^\pm_{\theta_0}$ implies analyticity in the corresponding strip in which Re$\theta$ is not constraint follows from the explicit formulas \fer{eqn6_16}, \fer{itheta} and \fer{A2.7}.)
The
analytic continuations of the operators (if they exist) are denoted by the same
symbols.

We define the family $K_\theta$ for $\theta \in \{ \theta \in
\BC^2 \big| |\rIm \theta| < \theta_0 \}$ by the explicit expressions
(\ref{eq6.18}), (\ref{eqn6_16}), (\ref{itheta}) and \fer{A2.7}. Clearly,
$D(\Lambda)\cap D(N) \subset D(L_{0\theta})$ and on this domain the family
$L_{0\theta}$ is manifestly strongly analytic in $\theta \in \{ \theta \in
\BC^2 \big| | {\rm Im} \theta | < \theta_0 \}$. It is shown in
Appendix \ref{App_Repr} that for $|{\rm Im}\theta| < \theta_0$ we have $D(\Lambda^{1/2})
\subset D(I_\theta)$ and $I_{\theta} f$ is analytic $\forall f \in
D(\Lambda^{1/2})$. Here Condition (AA) is used. Hence the family
$K_\theta$ for $\theta \in \{ \theta \in \BC^2 \big| |\rIm \theta|
< \theta_0 \}$ is bounded from $D(\Lambda) \cap D (N)$ to $\cH$ (and $K_\theta f$ is analytic in $\theta\in \{ \theta \in \BC^2 \big| |\rIm \theta|
< \theta_0 \}$, $\forall f\in\dom(\Lambda)\cap\dom(N)$).
Moreover, for $|{\rm Im}\ \theta| > 0$ the operators $K_\theta$ are
closed on the domain $D(\Lambda)\cap D(N)$.

However, $\{ K_\theta  |\, |{\rm Im}\theta|<\theta_0\}$  is {\it not} an analytic family in the sense of Kato. The problem here is the lack of coercivity -- the
perturbation $I$ is not bounded relatively to the unperturbed
operator $L_0$. To compensate for this we have chosen the
deformation $U_{\theta}$ in such a way that the operator
$M_{\theta} := {\rm Im} L_{0, \theta}$ is coercive for
$\rm{Im}\theta > 0$ , i.e., the perturbation $I_{\theta}$, as well
as ${\rm Re} L_{0, \theta}$, are bounded relative to this operator.
The problem here is that $M_{\theta} \to 0$ as $\rm{Im}\theta \to
0$ so we have to proceed carefully.

The next result is similar to one in \cite{BFS:RtE}, but the proof
given below is simpler than that of \cite{BFS:RtE}.

\begin{theorem} 
\label{thm_6.1}
Assume that Condition (AA) holds and let $\theta_0 = (\delta_0,
\tau_0)$ be as in that condition. Take an
\begin{equation}
a>\frac{g^2}{\sin({\rm Im}\delta)}C_0^2\left(\sum_{j=1,2}\|G_j\|_{1/2,\theta}\right)^2,
\label{condition on a}
\end{equation}
where 
\begin{equation}
C_0:=C (1+\beta_1^{-1/2}+\beta_2^{-1/2}),
\label{C_0}
\end{equation}
and where $C$ is a constant depending only on $\tan\delta_0$. Then we have:
\begin{enumerate}
\item[{\rm (i)}] $\left\{ z \in \BC | \rIm z \le -a \right\}\subseteq
\rho(K_{\theta})$ (the resolvent set of $K_\theta$) if $\theta\in S_{\theta_0}^+$; if in addition $K=K^*$ then we can take $\theta\in \overline{S_{\theta_0}^+}$;
\item [{\rm (ii)}] The family $K_{\theta}$ is analytic of
type A (in the sense of Kato) in $\theta\in S_{\theta_0}^+$;
\item [{\rm (iii)}] If $K=K^*$, then, for any $u$
and $v$ which are $U_{\theta}$-analytic in a strip $\left\{ \theta
\in \BC^2 | \, 0 \le {\rm Im} \theta < \theta_1 \right\}$,  for some
$\theta_1 = (\delta_1 ,\tau_0) ,\delta_1\in
[0,\min\{\pi/3,\theta_0\})$, the following relation holds:
\begin{equation}\label{eq6.21}
\left< u, (K-z)^{-1} v \right> = \left<
u_{\overline{\theta}}, (K_{\theta} - z)^{-1} v_{\theta}
\right>,
\end{equation}
where $u_{\theta} = U_{\theta} u$, etc., for $\rIm z
\le -a$ and $0 < \rIm\theta < \theta_1/2$.
\end{enumerate}

Similar statements hold also for $-\theta_0 < {\rm Im} \theta \le
0$.
\end{theorem}

{\it Proof.\ } 
\begin{enumerate}
\item[(i)] This statement is a special case of the following
proposition (estimate (\ref{eqn7_8}) below suffices). Let $C_{a,b}$ be the truncated wedge
\begin{eqnarray} 
\lefteqn{C_{a,b} :=  \label{eqn_7.0}}\\
&&\left\{ z\in \BC \ | \ \rIm z > -a/2, \ | \rRe z | <
2[(\sin b)^{-1}+a/4] (\rIm z +a) +\|L_p\|+1 \right\}.
\nonumber
\end{eqnarray}

\begin{prop}\label{prop7.2}
Let $\theta\in S_{\theta_0}^+$, and take  
$a$ as in \fer{condition on a}. Then $\sigma (K_{\theta}) \subset C_{a,\rIm \delta}$, and for $z\in \BC \backslash C_{a, \rIm \delta}$ we have 
\begin{equation}\label{eq6.3}
\| (K_\theta-z)^{-1}\| \le [\dist (z,C_{a, \rIm \delta})]^{-1}.
\end{equation}
\end{prop}

{\it Proof.\ }
To keep notation simple we prove the proposition for $\theta$
purely imaginary: $\theta = i\theta'$, $\theta' = (\delta', \tau')
\in \BR^2$. In this case the operator $M_{\theta}$ is of the form
\begin{equation}
M_{\theta} = \sin\delta'\Lambda + \tau' N.
\label{michael2}
\end{equation}
The proof below is based on the following bounds on the
interaction, which, to simplify the notation, we formulate for the
case $\theta = i \theta'$ only.

\begin{lemma} 
\label{lemma7_3}
Let $\mu$ be the same as in Condition (AA) above. We have 
\begin{eqnarray}
&& \norm{(M_\theta + a)^{-1/2} I_\theta (M_\theta + a)^{-1/2}}
   \le C_0 \frac{\sum_{j=1}^2 \norm{G_j}_{1/2, \theta}}{\sqrt{a \sin\delta'}}, \label{eqn_7_5} \\
&& \norm{\chi_{M_\theta \le \rho } I_\theta \chi_{M_\theta \le
\rho }}
  \le C_0 \left( \frac{2\rho}{\sin\delta'} \right)^{\!\mu} \sum_{j=1}^2
  \norm{G_j}_{\mu,\theta} , \label{eqn_7_6} \\
&& \left| \innerprod{\psi}{I_\theta \psi} \right|
  \le \frac{\varepsilon}{\sin\delta'} C_0^2\left(\sum_{j=1}^2\|G_j\|_{1/2,\theta}\right)^2 \innerprod{\psi}{M_\theta
  \psi} + \frac{1}{\varepsilon} \norm{\psi}^2,\ \ 
\label{eqn_7_7}
\end{eqnarray}
for any $a, \rho, \varepsilon >0$, and where $C_0$ is given in \fer{C_0}. 
Similar estimates hold also if we replace
$I_\theta$ by either ${\rm Re}I_\theta$ or ${\rm Im}I_\theta$.
\end{lemma}

This lemma follows from Proposition~\ref{relbounds} of
Appendix~\ref{App_RelBound} and equation \fer{aaa3} (cf. \cite{BFS:RG}). The norms on the
r.h.s. of (\ref{eqn_7_5}) - (\ref{eqn_7_7}) are defined in
(\ref{eqn_cond_B2}). 

Now we use the lemma above to prove Proposition \ref{prop7.2}.
First we determine the numerical range, NR$(K_\theta)$, of the operator
$K_{\theta}$. Let $u\in{\cal D}(M_\theta^{1/2})$ and $\|u\| = 1$.
Recall the notation $|A| := (A^{*}A)^{1/2}$ and remember that we
assumed that $\theta = (i\delta', i\tau')$. By estimate
(\ref{eqn_7_7}) and $|{\rm Re}L_{0,\theta}| \le \| L_p \| + \cos
\delta'\Lambda$ we have
\begin{equation} 
\label{eqn7_6}
  |{\rm Re}\left< K_\theta \right>_u| \le \left< \Lambda + \frac{C^2_1g^2}{\sin\delta'}M_\theta + \|L_p\|+ 1
  \right>_u,
\end{equation}
where $\langle A\rangle_u:=\scalprod{u}{Au}$, and we have set $C_1:=C_0\sum_{j=1}^2\|G_j\|_{1/2,\theta}$. Next, using that ${\rm Im}K_\theta = M_\theta + g{\rm
Im}I_\theta$, we write
\begin{equation*}
  {\rm Im}\left< K_\theta + ia \right>_u = \left<
  \overline{M_\theta}^{1/2} (1 +R) \overline{M_\theta}^{1/2} \right>_u,
\end{equation*}
where $\overline{M_\theta} = M_\theta+a$ and $R = g
\overline{M_\theta}^{-1/2 } {\rm Im}I_\theta
\overline{M_\theta}^{-1/2}$. Using estimate (\ref{eqn_7_5}) we
obtain $\|R\| \le \frac{g C_1}{\sqrt{a \sin\delta'}}$. Hence if
\begin{equation}
g C_1  < \frac 12 \sqrt{a\sin\delta'},
\label{a condition}
\end{equation}
then we have  
\begin{equation} 
\label{eqn7_8}
   {\rm Im} \left< K_\theta \right>_u  +a \ge \frac{1}{2}
  \left< \overline{M_\theta} \right>_u\geq a/2.
\end{equation}
This shows that ${\rm Im}\av{K_\theta}_u\geq -a/2$. It follows from $M_\theta = \sin\delta' \Lambda+\tau'N$ and \fer{eqn7_6} that 
$ |{\rm Re}\left< K_\theta \right>_u| \le \left<\frac{1+C^2_1g^2}{\sin\delta'}M_\theta + \|L_p\|+ 1
  \right>_u,
$
and hence, by $\av{M_\theta}_u \leq \av{\overline{M_\theta}}_u$ and \fer{eqn7_8}, 
\begin{equation}
  \left| {\rm Re}\left< K_\theta \right>_u \right| \le2 \frac{1+C_1^2g^2}{\sin\delta'}({\rm Im}\av{K_\theta}_u+a) +\|L_p\| +1.
\end{equation}
Using the bound \fer{a condition} in the last expression shows that 
NR$(K_\theta)\subset C_{a,\delta'}$, where $C_{a,\delta'}$ is the truncated wedge \fer{eqn_7.0}, provided condition \fer{a condition} is satisfied. In particluar, the spectrum of the operator $K_\theta$ is  inside $C_{a, \delta'}$. Moreover, for $z \notin
C_{a, \delta'}$ and $u$ as above we have the estimate
\begin{equation}
  \| (K_\theta - z)u \| \ge |\left< K_\theta \right>_u - z| \ge {\rm dist}(z, C_{a, \delta'}),
\end{equation}
which, by taking $u = (K_\theta - z)^{-1}v / \|(K_\theta - z)^{-1}v
\|$, implies (\ref{eq6.3}).
\item[(ii)] Estimates
$\|u\| \| (K_\theta - z ) u \| \ge \rIm \langle u , (K_\theta - z
) u \rangle $ and (\ref{eqn7_8}) imply for $\rIm z \le - a$:
\begin{equation}\label{9.16}
\| (K_\theta - z) u \| \ge \frac{\sqrt a}{2} \| M^{1/2}_\theta u  \| \ .
\end{equation}

The last estimate can be rewritten as
\begin{equation}\label{9.17}
\| M^{1/2}_\theta  (K_\theta - z)^{-1} \| \le \frac{2}{\sqrt a} \ .
\end{equation}
Similarly we have
\begin{equation}\label{9.18}
\| M^{-1/2}_\theta \partial_\theta K_\theta M^{-1/2}_\theta \| \le
C,
\end{equation}
where $\partial_\theta$ stands for $\partial_\delta$, $\partial_\tau$. 
The last two estimates and the computation
\begin{equation}\label{eq6.25}
\partial_\theta (K_\theta -z)^{-1} = - (K_\theta - z)^{-1}
\partial_\theta K_\theta (K_\theta - z)^{-1}
\end{equation}
imply that $(K_\theta - z)^{-1} $ is analytic in $\theta \in
S^\pm_{\theta_0} $, provided $\rIm z \leq -a$.

\item[(iii)] Now to fix ideas we assume that $\rIm \theta \geq 0$ and
$\rIm z < -a$. For $\alpha >0$ we define $K^{(\alpha)} := K + i\alpha N$.  Then
$K^{(\alpha)}_\theta:=U_\theta K^{(\alpha)} U^{-1}_\theta = K_\theta + i\alpha N$ and by standard estimates
similar to those in Proposition A.1 of Appendix A,
$(K^{(\alpha)}_\theta - z)^{-1}$ is analytic for $\rIm \theta > 0$, uniformly bounded  and strongly continuous for $\rIm \theta \ge 0$. (To prove the latter property it suffices to show that $(K_\theta^{(\alpha)}-z)^{-1}$ is strongly continuous on the dense set $\dom(\Lambda)$ which is straightforward.) 
Let $u$ and $v$ be
$U(\theta)$-analytic for $| \rIm \theta | < \delta_1$ for some
$\frac23 \theta_0 > \delta_1 > 0$.  Then in a standard way
\begin{equation}\label{9.19}
\langle u , (K^{(\alpha)} - z)^{-1} v \rangle = \langle u_{\thetabar },
(K_{\theta}^{(\alpha)} - z)^{-1} v_\theta \rangle
\end{equation}
for $\theta$ with $\rIm \theta>0$. Let now $v\in \cD (N)$ (then $v_\theta\in\dom(N)$).  With a help of the second resolvent equation
\begin{equation*}
(K^{(\alpha)}_{\theta} - z)^{-1} = (K_\theta - z)^{-1} -
(K_\theta^{(\alpha)} - z)^{-1} i\alpha N(K_\theta- z)^{-1},
\end{equation*}
we see that both sides of (\ref{9.19}) converge as $\alpha\to 0$, with
(\ref{eq6.21}) resulting in the limit. Finally, we remove the
constraint $v\in \cD(N)$ using a standard density argument. Namely, we approximate the $U_\theta$-analytic vectors $u$ and $v$ by the vectors $(1+\epsilon N)^{-1}u$ and $(1+\epsilon N)^{-1}v$ which belong to $\dom(N)$ and, since $U_\theta N U_\theta^{-1}=N$, are $U_\theta$-analytic as well. \hfill$\blacksquare$
\end{enumerate}

{\it Remark.} The other two complex deformations, \cite{JP:QFII}
and \cite{BFS:RtE}, are not suitable
technically in the present context due to the following reasons:
\begin{description}
\item{-} \cite{JP:QFII} leads to the problem in contour
integration for the resolvent representation of the dynamics (see
\cite{mms2}) 
\item{-} \cite{BFS:RtE} leads to a spectrum
in which an eigenvalue at $0$ is embedded at a ``tip'' of the
continuous spectrum and consequently it is technically more
difficult to define the pole approximation in this case (see
\cite{mms2}).
\end{description}

\section{Spectral Analysis of $K_{\theta}$} 
\label{Sect_7}

In this section we describe the spectrum of the operator $K_\theta$, ${\rm Im}\theta>0$,  in the half-space 
\begin{equation}
\label{eq6.7}
S = \left\{ z\in \BC \big| \rIm z <\frac{\sin (\rIm \delta )
}{4} \rho_0 \right\},
\end{equation}
where $0<{\rm Im}\delta<\delta_0$, and $\rho_0 \in (0,\sigma/2)$, c.f. \fer{gap}. 

Let $e$ be an eigenvalue of $L_p$ and let $\Lambda_e$ be the
operator acting on ${\rm Ran} \chi_{L_p = e}$ defined by
\begin{equation}
\Lambda_e := - P_e I (L_0-e+i0)^{-1} I P_e,
\label{m2}
\end{equation}
where $P_e = \chi_{L_p = e} \otimes \chi_{L_r = 0}$. Since ${\rm
Ran} (I P_e)$ is orthogonal to ${\rm Null}(L_0-e)$ this operator can
be, at least in principle, defined.  To show that it is well
defined we consider the operator $P_e I_\theta L^{-1}_{0\theta}
I_{\theta} P_e$ which {\it is} well-defined since ${\rm Ran}
(I_\theta P_e)$ is orthogonal to ${\rm Null} (L_{0\theta}-e)$ (and $e$ is an isolated eigenvalue of $L_{0\theta}$), is
independent of $\theta$ and is equal to $\Lambda_e$.
The operator $\Lambda_e$ is called the {\it level shift operator}.

The main result of this section is Theorem \ref{specthm}, which shows how the level shift operators $\Lambda_e$ determine the essential features of the spectrum of $K_\theta$.

{}For $\rho_0\in(0,\sigma/2)$ we decompose the half space $S$  into the strips
\begin{equation}
\label{eq6.8}
S_e = \left\{ z\in S |\, |\rRe z -e | \le \rho_0 \right\}
\end{equation}
where $e \in \sigma (L_p)$, and we set
$\oS = S \bs \bigcup\limits_{e\in \sigma (L_p) } S_e$, 
so that $S = \bigcup\limits_{e \in \sigma (L_p ) } S_e \cup \oS$.

\begin{theorem}
\label{specthm}
Assume condition (AA) holds. Take $0<|g|<\sqrt{\rho_0}\,  g_0$ (c.f. \fer{eq6.7}, \fer{g_0}), and $e\in\sigma(L_p)$. Let $\alpha=(\mu-1/2)/(\mu+1/2)$, where $\mu>1/2$ is given in Condition (AA).
\begin{itemize}
\item[1.] We have $\sigma (K_{\theta}) \subset \, \bigcup_{e\in\sigma(L_p)}S_e$.
\item[2.] 
Choose $\rho_0=|g|^{2-2\alpha}$. Suppose ${\rm Im}\Lambda_e:={\textstyle\frac{1}{2i}}(\Lambda_e-\Lambda_e^*)\geq\gamma_e>0$. If $|g|^\alpha\lless\gamma_e$, then 
\begin{equation}
\sigma(K_\theta)\cap S_e\subset \{z\in{\mathbb C}\,|\, {\rm Im\,}z\geq{\textstyle \frac 12} g^2\gamma_e\}.
\end{equation}
\item[3.]
Choose $\rho_0=|g|^{2-2\alpha}$. Suppose that $\Lambda_e$ has a simple eigenvalue $\lambda_e$, and that ${\rm Im}\,\Big(\sigma(\Lambda_e)\backslash\{\lambda_e\}\Big)\geq {\rm Im}\lambda_e+\delta_e$, for some $ \delta_e>0$. There is a $C>0$ s.t. if $0<|g|<Cg_2$, where 
\begin{equation}
g_2:=\min[( \delta_e)^{1/\alpha}, (\tau')^{\frac{1}{2+\alpha}}],
\label{g_2}
\end{equation}
then 
\begin{equation}
\sigma(K_\theta)\cap S_e \subset \{z_0\}\cup \{z\in {\mathbb C}\, |\, {\rm Im} z\geq g^2{\rm Im}\lambda_e+ \textstyle{\frac 12} \min(g^2\delta_e,\tau')  \},
\label{??}
\end{equation}
where $z_0$ is a simple isolated eigenvalue of $K_\theta$, satisfying $|z_0-e-g^2\lambda_e| =O(|g|^{2+\alpha})$. Moreover, $g\mapsto z_0(g)$ is analytic in an open complex neighbourhood of the set $0<|g|<\min[(g_0)^{1/\alpha},g_2]\subset{\mathbb R}$.

\end{itemize}
\end{theorem}

{\it Remark.\ } The analysis leading to Theorem 6.1 works also for infinite dimensional particle systems. We need $\dim{\cal H}_p<\infty$ in order to verify the assumptions $\gamma_e>0$, $\delta_e>0$, see Proposition \ref{absprop2} and Assumption (B), \fer{eqn_2.7}.



{\it Proof of Theorem \ref{specthm}.\ } 
1.\ We use the operator $M_\theta := {\rm Im} L_{0,\theta} >
0$ and the representation
\begin{equation}
  K_{\theta} - z = (M_{\theta} + a)^{1/2} (A + B) (M_{\theta} +
  a)^{1/2},
\label{7.12.}
\end{equation}
where $a = \frac{\sin\delta'}{2}\rho_0$, $A := (M_{\theta} +
a)^{-1/2}(L_{0,\theta} - z)(M_{\theta} + a)^{-1/2}$ and $B = g
(M_{\theta} + a)^{-1/2} I_{\theta} (M_{\theta} + a)^{-1/2}$. For
$z = x - iy \in \overline{S}$, the operator $A$ has a spectral gap
independent of the coupling constant $g$. Specifically, we claim
that
\begin{equation}\label{eqn7_20}
  \|Au\| \ge \frac{1}{10} \|u\|.
\end{equation}
To prove this claim we observe first that the operators
$M_{\theta}$ and $L_{0,\theta}$ commute and that $A$ is a normal
operator. Next, since ${\rm Im} L_{0,\theta} = M_{\theta}$ we have that
${\rm Im} A = (M_{\theta} + a)^{-1}(M_{\theta} + y)$. On the subspace
$\{M_\theta \ge 2a\}$ we have, for $z \in \overline{S}$ (thus $y>-a/2$), $M_\theta
+ y \ge \frac{1}{2}(M_\theta + a)$ and therefore $|A| \ge {\rm Im} A \ge
\frac{1}{2}$. On the subspace $\{M_\theta \le 2a\}$ we estimate
\begin{equation}
|A| \ge |{\rm Re} A| \ge (1/3a) |L_p + \cos \delta' L_r - x|\ .
\end{equation}
Now, recall that $\theta = (i\delta', i\tau')$ and use
 \fer{eqn6_16} to conclude that $M_{\theta} = \sin {\delta'}\Lambda +
\tau' N$. Hence $|L_r| \le (\sin{\delta'})^{-1}M_\theta \le
2a/\sin{\delta'}$. Since $L_{0,\theta}=L_p+\cos\delta' L_r+i\sin\delta'\Lambda+i\tau'N$, we have for $z=x-iy\in\overline S$
\begin{equation*}
  |A| \ge \frac{1}{3a} \min_{e \in \sigma(L_p)}\left\{
  |e-x|-|\cos\delta'L_r|\right\} \ge \frac{1}{3a} \left( \rho_0 - 2a \cos\delta' \right).
 \end{equation*}
Remembering that $a = \frac{\sin\delta'}{2}\rho_0$ we conclude
that on the subspace $\{M_\theta\leq 2a\}$
\begin{equation}
|A|  \ge  \frac{2}{3 \sin\delta'} \left(1-\cos\delta'\,\sin\delta'\right)
 \ge  \frac{2(\sqrt 2-1)}{3\sqrt 2\sin\delta'} >\frac{1}{10\sin\delta'}>1/10. 
\label{eqn_7_22}
\end{equation}
We used that $\sin\delta'\cos\delta'\leq 2^{-1/2}$. 
Consequently, $|A| \ge \frac{1}{10}$ which implies (\ref{eqn7_20}).

On the other hand, for $|g|<\sqrt{\rho_0}\, g_0$, the operator $B$ is
small (see Lemma~\ref{lemma7_3}):
\begin{equation}\|B\| \le C_0 |g| \frac{\max_{j}
\|G_j\|_{1/2, \theta}}{\sqrt{a\sin\delta'}} \le
C_0 |g| \frac{\max_{j} \|G_j\|_{1/2,
\theta}}{\sqrt{\rho_0}\sin\delta'} <1/20.
\end{equation}
Hence the operator $K_{\theta} - z\id$ is invertible for $z \in
\overline{S}$ and for $|g|<\sqrt{\rho_0} g_0$. This completes the proof of 1.

2.\ To analyze the spectrum of $K_\theta$ inside $S_e$ we use
Feshbach maps introduced in \cite{BFS:QED, BFS:RG}, and extended in \cite{BCFS:SmoothFeshbach}. We review
quickly the definitions and some properties of these maps
referring the reader to \cite{BFS:RG,BCFS:SmoothFeshbach} for more detail. For simplicity we present here the original version, \cite{BFS:QED, BFS:RG}, though the refined one, \cite{BCFS:SmoothFeshbach}, the smooth Feshbach map, is easier to use from a technical point of view.  Let $X$ be
a Banach space and $P$ be a projection on $X$. Define $\oP := \id
- P$ and let $H_\oP := \oP H \oP$ and $R_\oP (H) := \oP H_\oP^{-1}
\oP$ if $H_\oP$ is invertible on ${\rm Ran}\oP$. We define the
Feshbach map $F_P$ by 
\begin{equation}\label{eqn_7_23}
  F_P(H) := P \left(H - H R_\oP(H)H\right)P
\end{equation}
on the domain
\begin{eqnarray}
  \D(F_P) & = & \{ H : X \to X | H_\oP \text{ is invertible},
   \nonumber \\
  && \hphantom{\{ } {\rm Ran}P \subseteq \D(H) \text{ and }
  {\rm Ran}R_\oP(H) \subseteq \D(PH\oP) \}. \label{eqn_7_24}
\end{eqnarray}
A key property of the maps $F_P$ is given in the following
statement proven in \cite{BFS:RG}:

{\theorem[Isospectrality Theorem] \label{thm_7_4}
\begin{enumerate}
  \item[{\rm (i)}] $0 \in \sigma(H) \iff 0 \in \sigma(F_P(H))$,
  \item[{\rm (ii)}] $H \psi = 0 \iff F_P(H)\varphi = 0$
  with $\varphi = P\psi$ (``$\Rightarrow$") and $\psi=(\id -
  R_\oP(H)H)\varphi$ (``$\Leftarrow$").
\end{enumerate}}

Thus, Feshbach maps have certain isospectrality properties while
reducing operators from the original space $X$ to the smaller
space ${\rm Ran}P$.

Now, we use Feshbach maps $F_{P_{e\rho}}$ with projections
$P_{e\rho}$ defined as
\begin{equation} \label{eqn7_23}
  P_{e\rho} := \chi_{L_p=e} \otimes \chi_{M_{\theta} \le \rho}.
\end{equation}
Here, recall, $\chi_{L_p = e}$ is the eigenprojection for the
operator $L_p$ corresponding to an eigenvalue $e\in\sigma(L_p)$
and $\chi_{M_{\theta} \le \rho}$ is the spectral projection for the
self-adjoint operator $M_{\theta}$ corresponding to the spectral
interval $[0,\rho]$ (remember that $M_{\theta}$ is a positive
operator).

{\lemma
\label{lem_7_5}
Assume Condition (AA) holds. Let $|g|<\sqrt{\rho_0}\, g_0$. If $z \in S_e$ then $K_{\theta z} := K_\theta
- z \in \D(F_{P_{e\rho_0}})$, and the operator $K^{(1)}_{\theta z}
:= F_{P_{e\rho_0}}(K_{\theta z})$ acting on ${\rm Ran}P_{e\rho_0}$
is of the form
\begin{equation}\label{eqn_7_26}
  K^{(1)}_{\theta z} = (e-z) \id + L_{r\theta} + g^2 \Lambda_e +
  O(\epsilon (g, \rho_0)),
\end{equation}
where $\rho_0 \in (0,\sigma/2)$, the remainder is estimated in operator norm, and for any $|g|, \rho>0$, 
\begin{equation}
\epsilon (g, \rho) := |g|\rho^{\mu} +
|g|^{3}\rho^{-1/2} +|g|^{2}\rho^{2\mu - 1}.
\end{equation}
 }

We give here a short proof of Lemma \ref{lem_7_5}. Another proof is obtained by an easy translation of Theorem~V.6 and Lemma~V.9 of \cite{BFS:RtE}.

{\it Proof of Lemma \ref{lem_7_5}.\ } In this proof we write $\rho$ for
$\rho_0$. In order to prove that $K_{\theta z} \in \D
(F_{P_{e\rho}})$ we let $W := M_\theta + \rho$, note that $W$
commutes with $L_{0\theta}$, and write
\begin{equation} \label{eqn_7.27}
  \oP_{e\rho} K_{\theta z} \oP_{e\rho} = \oP_{e\rho} W^{1/2} [A +
  B] W^{1/2} \oP_{e\rho},
\end{equation}
where $A := W^{-1} (L_{0\theta}- z)$ and $B := gW^{-1/2} I_\theta
W^{-1/2}$. Using that the operator $A$ is normal and that its
spectrum on ${\rm Ran}\oP_{e\rho}$ can be explicitly found, one
can show easily that it is invertible on ${\rm Ran}\oP_{e\rho}$, 
and $\norm{A^{-1}} \le C$, uniformly in $\rho$ and $g$ (cf. the
proof of \fer{7.12.} given after that equation). By Lemma
\ref{lemma7_3},
\begin{equation}
\norm{B} \le C_0 |g| \frac{\max_j \|G_j\|_{1/2,
\theta}}{\sqrt{\rho\sin\delta'}}.
\end{equation}
Hence, for $|g|<\sqrt{\rho}\, g_0$, the operator $A+B$ is
invertible and therefore so is $\oP_{e\rho} K_{\theta z}
\oP_{e\rho}$ on ${\rm Ran}\oP_{e\rho}$. It is easy to see that the
other conditions in the definition of $D(F_{P_{e\rho}})$ (see
Eqn \fer{eqn_7_24}) are satisfied and therefore $K_{\theta z} \in
\D(F_{P_{e\rho}})$.

Next, in view of definition (\ref{eqn_7_23}) we compute
\begin{equation} \label{eqn_7.28}
  P_{e\rho} K_{\theta z} P_{e\rho} = (e-z) \id + L_{r\theta}
  + g P_{e\rho} I_\theta P_{e\rho}
\end{equation}
acting on ${\rm Ran} P_{e\rho}$. By (\ref{eqn_7_6}) and with $\mu$ as in Condition (AA)
\begin{equation} \label{eqn_7.29}
  g P_{e\rho} I_\theta P_{e\rho} = O(g\rho^\mu).
\end{equation}

Using (\ref{eqn_7.27}), expanding $\oP_{e\rho} (\oP_{e\rho}
K_{\theta z} \oP_{e\rho})^{-1} \oP_{e\rho}$ in the Neumann series in
$B$, and using that $\norm{B} \le C|g|\rho^{-1/2}$, we find
\begin{equation} \label{eqn_7.30}
-g^2 P_{e\rho} I_\theta R_{\oP_{e\rho}}(K_{\theta z}) I_\theta
P_{e\rho} = g^2 \Lambda_{e\rho\theta} + O(g^3\rho^{-1/2}),
\end{equation}
where $\Lambda_{e\rho\theta} := P_{e\rho} I_\theta \oP_{e\rho}
L_{0\theta}^{-1} \oP_{e\rho} I_\theta P_{e\rho}$.

To estimate the operator $\Lambda_{e\rho\theta}$ we use the
expression of $I_\theta$ in terms of creation and annihilation
operators, pull through the annihilation operators to the right
until they either contract or hit the projections $P_{e\rho}$, and use
estimates (\ref{relbnd}) and (\ref{eqn_A.3.6}) for $a_{j\ell,r}(k)
P_{e\rho}$ and $P_{e\rho} a^*_{j\ell,r}(k)$. As a result we obtain
\begin{equation}\label{eqn_7.31}
  \Lambda_{e\rho\theta} = \Lambda_e P_{e\rho} + O(\rho^{2\mu -1}),
\end{equation}
where $\Lambda_e$ acts nontrivially only on the particle Hilbert space 
(see Appendix \ref{appC'} for more detail). Using relations
(\ref{eqn_7.28}) -- (\ref{eqn_7.31}) in the expression for
$F_{P_{e\rho}}(K_{\theta z})$ (see (\ref{eqn_7_23})) we arrive at
(\ref{eqn_7_26}). 

This finishes the proof of Lemma \ref{lem_7_5}.\hfill$\blacksquare$ 

We now complete the proof of Theorem \ref{specthm}, parts 2 and 3. 
By the isospectrality of the map $F_{P_{e\rho_0}}$ and Lemma \ref{lem_7_5}, we have
\begin{equation}
\label{eq6.14}
\sigma (K_{\theta}) \cap S_e = \Big( \sigma  \left(L_{r\theta} + g^2
\Lambda_e  + O(\epsilon (g, \rho_0))\right)+ e \Big) \cap S_e.
\end{equation}

\begin{itemize}
\item[2.]  Since ${\rm Im} (L_{r\theta}+g^2\Lambda_e)\geq g^2\gamma_e$, and $\epsilon (g, \rho_0)\leq 3|g|^{2+\alpha}$, the numerical range of $L_{r\theta}+g^2\Lambda_e+O(\epsilon (g, \rho_0))$ is a subset of $\{{\rm Im}z\geq \frac 12 g^2\gamma_e\}$, provided $|g|^\alpha\lless\gamma_e$. The desired result follows from the fact that the spectrum of $L_{r\theta}+g^2\Lambda_e+O(\epsilon (g, \rho_0))$ is contained in the closure of the numerical range, and from \fer{eq6.14}.

\item[3.]

We start with the following result.
\begin{lemma}
\label{speclemma}
Let $A$ be a normal operator on a Hilbert space ${\cal H}_1$, and let $B$ be an operator on a Hilbert space ${\cal H}_2$, $\dim{\cal H}_2=d<\infty$. Then
\\
{\rm (i)} $\sigma(A\otimes\bbbone +\bbbone\otimes B) = \sigma(A) +\sigma(B)$,\\
{\rm (ii)} for $z\notin \sigma(A) +\sigma(B)$ we have 
\begin{equation}
\left\| (A\otimes\bbbone +\bbbone\otimes B-z)^{-1}\right\| \leq C\left[{\rm dist}(\sigma(A)+\sigma(B),z)\right]^{-n},
\label{npr10}
\end{equation}
where $1\leq n\leq d$ is the largest degree of nilpotency of the eigenvalues of $B$.\\
{\rm (iii)} Let $c$ be an isolated eigenvalue of $A\otimes\bbbone+\bbbone\otimes B$. There is a $p$, $1\leq p\leq d$, s.t. for $i=1,\dots,p$ we have $c=a_i+b_i$, where the $a_i$ are isolated eigenvalues of $A$ and the $b_i$ are eigenvalues of $B$. The (Riesz) projection onto $c$ is $\sum_{j=1}^p\chi_{A=a_j}\otimes\chi_{B= b_j}$, where $\chi_{A=a}$ and $\chi_{B=b}$ are the (Riesz) projections onto $a$ and $b$, respectively.
\end{lemma}
We prove part 3 of Theorem \ref{specthm} using Lemma \ref{speclemma} and refer to the end of this section for a proof of the lemma. We approximate the operator $\Lambda_e$ by a family of operators $\Lambda_e^{(\eta)}$, satisfying $\|\Lambda_e-\Lambda_e^{(\eta)}\|\leq \eta$, where $\eta>0$ is arbitrarily small, and where $\Lambda_e^{(\eta)}$ has semisimple spectrum, with a simple eigenvalue at $\lambda_e$, and with ${\rm Im}\,\Big(\sigma(\Lambda_e^{(\eta)})\backslash\{\lambda_e\}\Big)\geq{\rm Im}\lambda_e+ \delta_e$. A possible realization of $\Lambda_e^{(\eta)}$ is as follows. Let $\Lambda_e=\sum_j(D_j+N_j)$ be the Jordan decomposition of $\Lambda_e$, i.e., $D_j=\ell_j\bbbone$ (here the $\ell_j$ are the eigenvalues of $\Lambda_e$), $N_j^{m_j}=0$. Define
\begin{equation}
\Lambda_e^{(\eta)}:=\sum_j\left( D^{(\eta)}_j+N_j\right),
\label{jordan}
\end{equation}
where (for $\ell_j$ non-semisimple) $D_j^{(\eta)}:={\rm diag}(\ell_j,\ell_{j,1}(\eta),\ldots,\ell_{j,m_j-1}(\eta))$, and where the $\ell_{j,k}(\eta)$ are arbitrary distinct complex numbers with imaginary part $\geq {\rm Im}\lambda_e+ \delta_e$, satisfying $|\ell_j-\ell_{j,k}(\eta)|\leq \eta$. 

Choosing $A=L_{r\theta}$, $B=g^2\Lambda_e^{(\eta)}$, we see from Lemma \ref{speclemma} (i), (iii) that 
the operator $L_{r\theta}+g^2\Lambda_e^{(\eta)}$ has a simple eigenvalue at $g^2\lambda_e$ and the rest of the spectrum is located in $\{z\in{\mathbb C}\,|\, {\rm Im\,}z\geq g^2{\rm Im}\lambda_e+\min(g^2 \delta_e,\tau')\}$. 

We use relation \fer{eq6.14} to investigate the spectrum of $K_\theta$ inside $S_e$. The error term in \fer{eq6.14} satisfies $O(\epsilon(g,\rho_0))=O(|g|^{2+\alpha})$.
 From \fer{npr10} (with $n=1$) and an elementary Neumann series estimate it follows that the spectrum of  $L_{r\theta}+g^2\Lambda_e+ O(\epsilon(g,\rho_0))$ lies in a neighbourhood of order $O(|g|^{2+\alpha}+g^2\|\Lambda_e^{(\eta)}-\Lambda_e\|)=O(|g|^{2+\alpha})$ of the spectrum of $L_{r\theta}+g^2\Lambda_e^{(\eta)}$ (for $\eta$ small enough). 
Moreover, since by our assumptions 
\begin{equation}
|g|^{2+\alpha}\lless \min(g^2 \delta_e,\tau')
\label{npr15}
\end{equation}
(see \fer{g_2}), one easily proves, using Riesz projections, that $L_{r\theta}+g^2\Lambda_e+ O(\epsilon(g,\rho_0))$ has a simple eigenvalue $z_0$ in an $O(|g|^{2+\alpha})$-neighbourhood of $g^2\lambda_e$. The rest of the spectrum of $L_{r\theta}+g^2\Lambda_e+ O(\epsilon(g,\rho_0))$ is located in $\{z\in {\mathbb C}\, |\, {\rm Im\,}z > g^2{\rm Im}\lambda_e+\frac12\min(g^2 \delta_e,\tau')\}$. The result \fer{??} follows from the isospectrality, \fer{eq6.14}. 

Fix an arbitrary $g'$, $0<|g'|<\min[(g_0)^{1/\alpha}, g_2]$. By the Kato-Rellich Theorem, $g\mapsto z_0(g)$ is analytic in a complex neighbourhood of $g'$. This completes the proof of Theorem \ref{specthm}, point 3, and hence the entire proof of Theorem \ref{specthm}. \hfill $\blacksquare$

{\it Proof of Lemma \ref{speclemma}.\ } By using the spectral representation of $A$ and the normal form of the operator $B$, \cite{kato} I.5.3, one obtains
\begin{equation}
(A\otimes\bbbone +\bbbone\otimes B-z)^{-1} =\sum_j\sum_{n=0}^{m_j-1}(-1)^n(A+b_j-z)^{-n-1}\otimes Q_j^{(n)},
\label{npr11}
\end{equation}
where $b_j$ are the eigenvalues of $B$, $Q_j^{(0)}=\chi_{B=b_j}$ is the projection (Riesz integral) onto the eigenvalue $b_j$, and, for $n\geq 1$, $Q_j^{(n)}=N_j^n$, with $N_j=Q_j^{(0)}N_j=N_jQ^{(0)}_j$ a nilpotent matrix, $N_j^{m_j}=0$. Assertions (i), (ii) follow.

Let $C$ be a circle of radius $r<\dist[c,(\sigma(A)+\sigma(B))\backslash\{c\}]$ around $c$. From \fer{npr11},
\begin{eqnarray}
\lefteqn{
\frac{1}{2\pi i}\oint_C dz (A\otimes\bbbone +\bbbone\otimes B-z)^{-1}  = \frac{1}{2\pi i}\oint_C dz \sum_j\sum_{n=0}^{m_j-1}(-1)^n }\nonumber\\
&&\times \left[(c-z)^{-n-1} \chi_{A=a_j}\otimes Q_j^{(n)} + (A+b_j-z)^{-n-1}(1-\chi_{A=a_j})\otimes Q_j^{(n)}\right].\ \ \ \ \ \ \ 
\label{npr11'}
\end{eqnarray}
The first term on the r.h.s. of \fer{npr11'} contributes only for $n=0$ (for each $j$ fixed), while the second term does not contribute at all. This concludes the proof of Lemma \ref{speclemma}. \hfill $\blacksquare$ 
\end{itemize}

\section{Absence of $\beta_1\beta_2$-normal stationary states}
\label{Sect_13}

In this section we prove Theorem~\ref{thm2.1}. Let $L = L_0 + g\pi(v) - g\pi'(v)$ be the standard
(self-adjoint) Liouville operator, \fer{standardliouvillian},  and let $L_\theta$ be its $U_\theta$-deformation. If Condition (C) is satisfied then the operator $\Lambda_0=i\Gamma_0$ is anti-selfadjoint, with $\Gamma_0\geq 0$ (see also Proposition \ref{absprop2} below, and \cite{BFS:RtE}). Let $\gamma_0\geq 0$ be the lowest eigenvalue of $\Gamma_0$, and let $\delta_0>0$ denote the distance of $\gamma_0$ to the rest of the spectum of $\Gamma_0$.

\begin{theorem}
\label{abs1thm}
Assume that conditions (A), (B) and (C) are obeyed for some
$0<\beta_1,\beta_2<\infty$, $\mu>1/2$, and set $\alpha=(\mu-1/2)/(\mu+1/2)$.
Assume $\gamma_0>0$.
 There is a constant $C>0$ s.t. if $0<|g|<Cg_3$, where 
\begin{equation}
g_3:=\min\big( (g_0)^{1/\alpha}, (\delta_0)^{1/\alpha},[\min(T_1,T_2)]^{\frac{1}{2+\alpha}}\big),
\label{npr20}
\end{equation}
then $L_\theta$ has a simple isolated eigenvalue $z_0(g)\in S_0$, satisfying $|z_0(g)-ig^2\gamma_0|=O(|g|^{2+\alpha})$, and the rest of the spectrum of $L_\theta$ inside $S_0$ lies in the region $\{z\in{\mathbb C}\,|\, {\rm Im\,}z\geq \frac12\min(g^2\delta_0,\tau')\}$.

Moreover, we have ${\rm Im\,}z_0(g)>0$, for all $0<|g|<Cg_3$, except possibly for finitely many values of $g$ in $\{C'(\gamma_0)^{1/\alpha}<|g|< C g_3\}$, for some constant $C'>0$.
\end{theorem}

{\it Remark.\ } The assertion $|z_0(g)-ig^2\gamma_0|=O(|g|^{2+\alpha})$ of the first part of Theorem \ref{abs1thm} shows that ${\rm Im}z_0(g)>0$ provided $|g|^\alpha\lless \gamma_0$. However, $\gamma_0$ depends on the difference of the reservoir temperatures, and it vanishes when both reservoirs are at the same temperature (see also the proof of Proposition \ref{absprop2}), and thus, $|g|^\alpha\lless\gamma_0$ is a too restrictive condition. The second part of Theorem \ref{abs1thm} resolves this difficulty, yielding a result for values of the coupling parameter $g$ uniform in the temperature difference of the reservoirs. \\

{\it Proof of Theorem \ref{abs1thm}.\ } We apply Theorem \ref{specthm}, part 3, with $e=0$. We have $\lambda_0=i\gamma_0$ and $\tau'=c\min(T_1,T_2)$ for some $c>0$, see after \fer{eq_2_11.1}, so the conditions $0<|g|<\sqrt{\rho_0}g_0$ and $0<|g|<Cg_2$ of Theorem \ref{specthm}, part 3, reduce to $0<|g|<Cg_3$. 

We must have ${\rm Im}\, z_0(g)\geq 0$, for otherwise, the selfadjoint operator $L$ would have an eigenvalue in the lower complex plane. 

To complete the proof of Theorem \ref{abs1thm} it remains to show that ${\rm Im}z_0(g)>0$, for all $0<|g|< g_3$, except possibly for a discrete set of values. Let $J$ be the open interval $J=\,]0,g_3[$. For any $g\in J$ there exists a complex disc $B(g)$ centered at $g$, s.t. $z_0(g)$ is analytic for $g\in B(g)$ (see also the proof of Theorem \ref{specthm}, part 3). Suppose that there is a sequence $g_n\rightarrow g'$, s.t. $g_n,g'\in J$, and s.t. ${\rm Im\,}z_0(g_n)={\rm Im\,}z_0(g')=0$.  By expanding $z_0(g)$ in a Taylor series around $g'$ it is readily seen that ${\rm Im\,}z_0(g)=0$ for all $g\in B(g')\cap J$. Given any closed interval $J_1\subset J$ one easily sees that $\inf_{g\in J_1} |B(g)|>0$, where $|B(g)|$ is the radius of the disc $B(g)$. Therefore, again by Taylor series expansion, it follows that ${\rm Im\,}z_0(g)=0$ for all $g\in J_1$. 

However, Theorem \ref{specthm}, part 2, shows that there is a $C'>0$ s.t. if $0<|g|<C'(\gamma_0)^{1/\alpha}$, then we have ${\rm Im\,}z_0(g)\geq\frac 12 g^2\gamma_0>0$. Consequently there cannot exist any accumulation point $g'$ inside $J$. The only possible such accumulation point is thus $g'=0$ or $g'=g_3$. The former is ruled out again due to  Theorem \ref{specthm}, part 2. By choosing a possibly smaller value of the constant $C$ we achieve that ${\rm Im\,}z_0(g)>0$, except possibly for finitely many values of $g$ in $\{C'(\gamma_0)^{1/\alpha}<|g|< Cg_3\}$.
\hfill $\blacksquare$

\begin{proposition}
\label{absprop2}
Assume Conditions (B), (C). Then 

{\rm (a)} $\gamma_0\geq C\min_j(\gamma_{0j})\frac{|\delta\beta|^2}{1+|\delta\beta|^2}$, where $\delta\beta=|\beta_2-\beta_1|$, $C>0$ is independent of $\beta_1,\beta_2$, and where $\gamma_{0j}$ are the constants given in \fer{eqn_2.7}.

{\rm (b)} There is a constant $c'>0$ s.t. if $\delta\beta<c'$ and $\|G_1-G_2\|<c'$ (see \fer{intop}), then $\delta_0\geq \gamma_{01}$. 
\end{proposition}

{\it Proof.\ }
Condition (C) ensures that the level shift operator  $\Lambda_0 : {\rm Ran}
\chi_{L_p = 0} \to {\rm Ran}\chi_{L_p = 0}$ is given by the
expression
$
\Lambda_0  := \sum_{j=1}^2 \Lambda_{0j}
$
with the operators $\Lambda_{0j} = i{\rm Im}\Lambda_{0j}=:i\Gamma_{0j}$ given  as in (\ref{m2})
with $e=0$, and with $I$ replaced by $I_j=\pi(v_j)-\pi'(v_j)$, see also \fer{intop}, \cite{BFS:RtE}. Moreover, we know from \cite{BFS:RtE} that $\Gamma_{0j} \ge 0$, that 
$\Gamma_{0j}$ has a simple eigenvalue at $0$ with eigenvector
$\Omega^p_{\beta_j}$, and that on the complement of ${\mathbb C}\Omega_{\beta_j}^p$, $\Gamma_{0j}\geq\gamma_{0j}$. By Condition (B), $\Gamma_{0j}>0$. Consequently, for $\beta_1 \not= \beta_2$,
$
  \Gamma_0 := \sum_{j=1}^2 \Gamma_{0j} >0
$.

(a) By analyzing the explicit form of the level shift operators, it is easy to show that $\Gamma_0 \ge C\min_j(\gamma_{0j})
\frac{|\delta\beta|^2}{1+|\delta\beta|^2}$. (In fact, $\Gamma_0\geq C\min_j(\gamma_{0j})(\delta\beta)^2[1-Z(\beta_1+\beta_2)/Z(\beta_1/2+\beta_2/2)]$, where $Z(\beta)={\rm Tr}(e^{-\beta H_p})$.)

(b) We view the gap $\delta_0$ as a function of the inverse temperatures $\beta_{1,2}$ and of the coupling operators $G_{1,2}$. Then we have $\delta_0(\beta_1=\beta_2,G_1=G_2)=2\gamma_{01}$. The result follows from the continuity of the operator $\Lambda_0$ in $G_j$ and $\beta_j$. \hfill $\blacksquare$\\

{\it Proof of Theorem \ref{thm2.1}.\ } 1. The conditions on $g$, $\delta\beta$, $\|G_1-G_2\|$ in Theorem \ref{thm2.1}, part 1, and Proposition \ref{absprop2}, (b), imply that Theorem \ref{abs1thm} is applicable. The latter theorem shows that $\sigma(L_\theta)\cap {\mathbb R}\cap S_0=\emptyset$. Hence the spectrum of non-deformed standard Liouville operator $L$, inside ${\mathbb R}\cap S_0$, is purely absolutely continuous. The result follows from Theorem \ref{ninvspecthm}. 

2. In the same way as for 1, combine Proposition \ref{absprop2}, (a), Theorem \ref{specthm}, part 2 (for $e=0$), and Theorem \ref{ninvspecthm}. \\

{\it Removing the high temperature condition $|g|\lless [\min(T_1,T_2)]^{\frac{1}{2+\alpha}}$ in \fer{g_1}, \rm \cite{mms3}.\ } The origin of this condition lies in Theorem \ref{abs1thm}, where we use the  bound
$$
O(\epsilon(g,\rho_0)) = O(|g|^{2+\alpha})\lless \min(g^2\delta_0,\tau')
$$
(see also \fer{npr20}) in order to be able to trace the simple isolated eigenvalue $z_0$ (c.f. \fer{npr15}, in the setting of Theorem \ref{abs1thm}, where $|g|^{2+\alpha}$ represents the error term $O(\epsilon(g,\rho_0))$ in \fer{eqn_7_26}). If this condition fails then we use the Feshbach map iteratively until the error term in the 
equation for the final iteration (corresponding to (\ref{eqn_7_26}) in the above case) is $\ll
\tau'\approx \min(T_1,T_2)$. Applying Theorems~V.17 and V.18 of \cite{BFS:RtE} we
conclude that the spectrum of the operator $L_\theta$ inside
$S_0$, $\rIm\theta
> 0$,  consists of a simple isolated eigenvalue at some point
$z_0$ with the rest of the spectrum lying in the half space
$
  \set{z \in \BC\ }{\ \rIm z \ge \rIm z_0 +\tau'/2}
$. The arguments in the proof of Theorem \ref{abs1thm} then show that $L_\theta$ does not have any real eigenvalues inside $S_0$, for all $0<|g|<C\min((g_0)^{1/\alpha}, (\delta_0)^{1/\alpha})$, except possibly for finitely many values of $g$.\\


{\bf Acknowledgements.} The authors are grateful to V. Bach, G.
Elliott, J. Fr\"ohlich, V. Jak\u si\'c and especially C.-A. Pillet for useful
discussions. 
Part of this work was done
while the first author was visiting the University of Toronto, 
the third author ETH Z\"urich, and the second and third authors ESI Vienna.  During work on this paper the
second author was at the University of Toronto on a DAAD
fellowship. The authors are grateful to these places for
hospitality.

\appendix
\section{Proof of existence of dynamics}
\label{appA} 
In this appendix we 
prove existence of the dynamics \fer{dyn1}. Recall the 
definition of  the
operator $L^{(\ell)}:=L_0+g\pi(v)$ 
and of the one parameter group $\sigma^{t}(B):=e^{itL^{(\ell)}} B 
e^{-itL^{(\ell)}}$,  $B \in \pi(\cA)''$. 
\begin{proposition}
\label{propa1}
Assume the operators $v_n \in \cA$ satisfy \fer{eq2.13}. Then
the 
integrands on the r.h.s. of \fer{dyn1} are continuous functions,
the series is absolutely convergent, the limit exists and equals
\begin{equation}
\psi^t(A)= {\rm Tr} (\rho \sigma^{t}(\pi(A)))
\label{michael1}
\end{equation}
and, consequently, is independent of the approximating operators. 
\end{proposition}

{\bf Proof.\ } Let $v_n \in \cA$ be an approximating
sequence for the operator $v$ satisfying \fer{eq2.13}. We define the selfadjoint operators $L^{(\ell)}_n:=L_0+g\pi(v_n)$ on the dense domain ${\cal
D}(L_0)$.  Let the one parameter group $\sigma^{t}_{(n)}$ on $\pi({\cal A})$ be given by
\begin{equation} \label{fm4}
\sigma^{t}_{(n)}(B):=e^{itL^{(\ell)}_n} B e^{-itL^{(\ell)}_n}.
\end{equation}
Set $ \sigma_0^{t}(\pi(A)):=\pi(\alpha_0^t(A))$ and let
$\psi$ be an $\omega_0$-normal state on 
$\cA$, i.e.
\begin{equation}
\psi(A)= {\rm Tr} (\rho  \pi(A))
\end{equation}  
for some positive, trace class operator $\rho$ on $\cH$ of trace 1. Then
using the definition $V_n = \pi(v_n)$ we find
\begin{equation} 
\label{dyn}
\psi([\alpha_0^{t_m}(v_{n}),\cdots 
[\alpha_0^{t_1}(v_{n}),\alpha_0^{t}(A)]\cdots]) = {\rm Tr} (\rho
[\sigma_0^{t_m}( V_n),\cdots [\sigma_0^{t_1}( V_n), \sigma_0^{t}(A)]\cdots]).
\end{equation}
Clearly the r.h.s. is continuous in $t_{1},\cdots, t_{m}$ and
therefore the integrals in \fer{dyn1} are well defined and, by a
standard estimate, the series on the r.h.s. of \fer{dyn1} converges
absolutely. In fact, using the Araki-Dyson series 
\begin{eqnarray} \label{mm32}
\sigma^{t}_{(n)}(\pi(A))&=&\sum_{m=0}^\infty(ig)^m\int_0^tdt_1\cdots
\int_0^{t_{m-1}}
dt_m  \ [\sigma_0^{t_m}(\pi(v_n)),
\cdots\nonumber\\
&& [\sigma_0^{t_1}(\pi(v_n)),\sigma_0^{t}(\pi(A))]\cdots],
\end{eqnarray} 
one can easily see that this series is nothing but the Araki-Dyson
expansion of the function ${\rm Tr} (\rho \sigma^{t}_{(n)}(\pi(A)))$. Thus
we have shown that the r.h.s. of \fer{dyn1} is equal to $\lim_{n
\rightarrow \infty} {\rm Tr} (\rho \sigma^{t}_{(n)}(\pi(A)))$.

Now, $V_n$ converges to $V$ strongly on the dense set ${\rm Span}
\{\pi(B\otimes W_1(f_1)\otimes W_2(f_2))\Omega_0|B\in{\cal B}({\cal H}_0), f_{1,2}\in L^{2}_0\}$ as follows from \fer{eq2.13} and the
relation
\begin{equation}
\|(V_n - V)\pi(A)\Omega_{0}\|^2 = \omega_{0}(A^{*}(v_{n}^* -
v^*)(v_n - v)A).
\end{equation}
Hence $L^{(\ell)}_n$ converges to $L^{(\ell)}$ strongly on the same
set. Since this set is a core for $L^{(\ell)}_n$ and $L^{(\ell)}$
we conclude that $L^{(\ell)}_n$ converge to $L^{(\ell)}$ in the
strong resolvent sense as $n \rightarrow \infty$ (\cite{RSI}, Theorem VIII.25), and therefore, $e^{itL_n^{(\ell)}}\rightarrow e^{itL^{(\ell)}}$ strongly.  Hence the
functions ${\rm Tr} (\rho \sigma^{t}_{(n)}(\pi(A))$ converge to ${\rm Tr} (\rho
\sigma^{t}(\pi(A)))$ which, in particular, shows \fer{michael1}. \hfill $\qed$

\section{Positive Temperature Representation and Relative Bounds}
\label{App_Repr}

\subsection{Jak\u{s}i\'{c}-Pillet Gluing} \label{AppC}

\renewcommand{\GL}{U}
In this appendix, we represent the Hilbert space ${\cal H}$ in a
form which is well suited for a definition of the translation
transformation. This representation is due to \cite{JP:QFII}. 

Consider the Fock space
\begin{equation} \label{eqn_A.21}
\cF:= \cF(L^2(X\times\{1,2\})),\ \ \ X= \BR\times S^2
\end{equation}
and denote $x=(u,\sigma)\in X$. 
The vacuum in $\cF$ is denoted by $\tilde{\Omega}_r$. The smeared-out
creation operator $a^*(F)$, $F\in L^2(X\times\{1,2\})$ is given by
\begin{equation*}
a^*(F)=\sum_\alpha\int_X F(x,\alpha) a^*(x,\alpha)
\end{equation*}
and analogously for annihilation operators. The CCR read
\begin{equation*}
[a(x,\alpha),a^*(x',\alpha')]=\delta_{\alpha,\alpha'}\delta(x-x').
\end{equation*}
Following \cite{JP:QFII}, we introduce the unitary map
\begin{equation}\label{eqn_A.20}
  \GL : \left[ \cF(L^2(\R^3)) \otimes \cF(L^2(\R^3)) \right]
    \otimes \left[ \cF(L^2(\R^3)) \otimes \cF(L^2(\R^3)) \right]
    \to \cF(L^2(X \times \{1,2\}))
\end{equation}
defined by
\begin{equation}
  \GL \left(\left[\Omega_{r1} \otimes \Omega_{r1}\right] \otimes
      \left[\Omega_{r2} \otimes \Omega_{r2}\right] \right) := 
\tilde{\Omega}_{r}
\end{equation}
and
\begin{eqnarray}
  \GL \Big( \left[a^*(f_1) \otimes \id + \id \otimes a^*(g_1)\right]
      \otimes \id \otimes \id  \hphantom{+ \GL^{-1}\Big)} && \nonumber \\
      + \id \otimes \id \otimes
      \left[a^*(f_2) \otimes \id + \id \otimes a^*(g_2)\right]
  \Big) \GL^{-1} & :=  & a^*(f\oplus g),
\end{eqnarray}
where, for $x=(u,\sigma)\in X$,
\begin{equation}
  \left[ f\oplus g \right](u, \sigma, \alpha) :=
  \begin{cases}
    u \, f_\alpha(u\sigma), & u \ge 0, \\
    u \, g_\alpha(-u\sigma), & u < 0.
  \end{cases}
\end{equation}
This map is extended to the Hilbert space $\cH = \cH^p \otimes
\cF$ in the obvious way. We keep the same notation for its
extension. \\
\indent The operators $L_{r1} \otimes \id_{r2} + \id_{r1} \otimes
L_{r2}$ and $N_{r1}\otimes \id_{r2} + \id_{r1} \otimes N_{r2}$ are
mapped under $\GL$ to the (total) free field Liouvillian and
number operator given by
\begin{eqnarray*}
L_f&=&\d\Gamma(u)=\sum_\alpha\int_X a^*(x,\alpha) u a(x,\alpha),\\
N&=&\d\Gamma(\bbbone)=\sum_\alpha\int_X a^*(x,\alpha) a(x,\alpha).
\end{eqnarray*}
Moreover, the interaction $I$ in the operator $K$ takes the form (c.f. \fer{K})
\begin{equation} \label{eqn_A_2_6}
\GL I \GL^{-1} = a^*(F_1) +a(F_2)
\end{equation}
where the $F_{j}\in L^2(X\times\{1,2\}, \cB(\cH_p\otimes\cH_p))$ are
explicitly given by ($x = (u,\sigma) \in X = \R \times S^2$)
\begin{eqnarray}
\lefteqn{
F_1(u,\sigma, \alpha)=}\label{eqn_def_F1}\\
&&\sqrt{\frac{u}{1-e^{-\beta_\alpha
u}}}\ |u|^{1/2}
\left\{
\begin{array}{ll}
G_{\alpha 1}(u\sigma)\otimes \bbbone_p -e^{-\beta_\alpha u/2}\bbbone_p\otimes \overline{G_{\alpha 4}}^*(u\sigma), & u>0\\
- G_{\alpha 2}^*(-u\sigma)\otimes\bbbone_p +e^{-\beta_\alpha u/2}\bbbone_p\otimes \overline{G_{\alpha 3}}(-u\sigma), & u<0
\end{array}
\right. 
\nonumber
\end{eqnarray}
\begin{eqnarray}
\lefteqn{
F_2(u,\sigma, \alpha)=}\label{eqn_def_F2}\\
&&\sqrt{\frac{u}{1-e^{-\beta_\alpha
u}}}\ |u|^{1/2}
\left\{
\begin{array}{ll}
G_{\alpha 2}(u\sigma)\otimes \bbbone_p -e^{-\beta_\alpha u/2}\bbbone_p\otimes \overline{G_{\alpha 3}}^*(u\sigma), & u>0\\
- G_{\alpha 1}^*(-u\sigma)\otimes\bbbone_p+e^{-\beta_\alpha u/2}\bbbone_p\otimes \overline{G_{\alpha 4}}(-u\sigma), & u<0
\end{array}
\right. 
\nonumber
\end{eqnarray}
Thus the operator $\tilde{K} := \GL K \GL^{-1}$ can be
written as
\begin{equation*}
  \tilde{K} = \tilde{L}_0 + g\tilde{I}
\end{equation*}
where $\tilde{I} = \GL I \GL^{-1}$ is given in (\ref{eqn_A_2_6})
and $\tilde{L}_0 := \GL L_0 \GL^{-1}$ is of the form
\begin{equation*}
  \tilde{L}_0 = L_p \otimes \id_f + \id_p \otimes L_f.
\end{equation*}

\subsection{Complex Deformation}
\label{b2}

Now we express the complex deformation operators $U_\theta$
introduced in Section~\ref{Sect6} in the Jak\u si\'c-Pillet glued
Hilbert space. For a function $F \in L^2\left( X \times
\{1,2\}\right)$ and $\theta = (\delta, \tau)$, $x=(u,\sigma)\in X$, define
\begin{equation} 
\label{eqn_complex_def}
  \left[\tilde{u}_\theta F \right](u, \sigma,\alpha) = e^{\frac{1}{2}\delta{\rm sgn}(u)}
F(j_\theta(u), \sigma, \alpha),
\end{equation}
where
\begin{equation} \label{eqn_A.25}
 j_\theta(u) =  e^{\delta{\rm sgn}(u)} u + \tau,
\end{equation}
and $\rm sgn$ is the sign function, $\mbox{sgn}(u)=1$ if $u\geq
0$, $\mbox{sgn}(-u)=-\mbox{sgn}(u)$. Next, we lift the operator
family $\tilde{u}_\theta$ from $L^2(X \times \{1,2\})$ to the
operator family, $\tilde{U}_\theta$, on ${\cal
H}^p\otimes\cF(L^2(X\times\{1,2\}))$ in a standard way (cf.
(\ref{eqn_6_3})). The family $\tilde{U}_\theta$ is related to the
family $U_\theta$ introduced in Section~\ref{Sect6} as
\begin{equation*}
  U_\theta = \GL \tilde{U}_\theta \GL^{-1}.
\end{equation*}

The operator $\tilde{K}$ becomes after spectral deformation
\begin{equation}
\tilde{K}_\theta:= \tilde{U}_\theta K \tilde{U}_\theta^{-1}
=\tilde{L}_{0,\theta}+g \tilde{I}_\theta \label{ktheta}
\end{equation}
where
\begin{eqnarray}
\tilde{L}_{0,\theta}
&=&L_p+\cosh\delta \ L_f+\sinh\delta \ \Lambda_f +\tau N, \label{m9}\\
\Lambda &=&\d\Gamma(|u|)=\sum_\alpha\int_X
a^*(x,\alpha)|u|a(x,\alpha),
\nonumber\\
\tilde{I}_\theta &=& a^*(F_{1,\theta})+a(F_{2,\theta}) \mbox{\ \ \
\ with \ \ $F_{j,\theta} = \tilde{u}_\theta F_j$.} \label{itheta}
\end{eqnarray}

This spectral deformation can be translated to the original space
$\cH$ as
\begin{equation}
\label{A2.6} K_\theta:=U^{-1} \tilde K_\theta U^{-1} =
L_{0,\theta} + g I_\theta
\end{equation}
where $L_{0,\theta}:=U^{-1} \tilde L_{0,\theta} U$ is given by 
\fer{eqn6_16} and
\begin{equation}
\label{A2.7} 
I_\theta = U^{-1} \tilde I_\theta U.
\end{equation}

\subsection{Relative Bounds} \label{App_RelBound}

We prove the bounds which imply Lemma~\ref{lemma7_3}. We will from
now on fix $\delta=i\delta'$ and $\tau=i\tau'$ for some
$\delta', \tau'>0$. Recall that the operator $M_\theta$ is given by
\begin{equation*}
M_\theta:=\rIm \tilde{L}_{0,\theta}=\sin\delta' \Lambda +\tau'
N\geq 0.
\end{equation*}

\begin{proposition}
\label{relbounds} 
For a function $F: X \times \{1,2\} \to {\cal
B}({\cal H}_p \otimes{\cal H}_p)$ set $F_\theta(x,\alpha)= e^{\rm{sgn}(u)\delta/2}
F(j_\theta(u),\sigma,\alpha)$, where $x=(u,\sigma)$ and $j_\theta(u)$ is given in
(\ref{eqn_A.25}), with $\theta = (i\delta', i\tau')$,
$\delta',\tau'>0$. Suppose that the function $F$ satisfies
\begin{equation} \label{ir}
 ||F||_\rho :=\left( \sum_\alpha \int\limits_{\sin(\delta') |u| + \tau'  \le \rho}
  \frac{\| F_\theta (x, \alpha) \|^2}{|j_\theta(u)|} \, dud\sigma\right)^{1/2}
  <\infty
\end{equation}
for some $0 < \rho \leq\infty$. Then we have the bounds
\begin{eqnarray}
\|a(F_\theta)M^{-1/2}_\theta\| &\leq& \frac{1}{\sqrt{\sin\delta'}}
||F||_\infty, 
\label{relbnd''}\\
\|a^*(F_\theta)M^{-1/2}_\theta\| &\leq& \|F_\theta\|_{L^2} +
\frac{1}{\sqrt{\sin\delta'}} \|F\|_\infty    
\label{A3.3}\\
\|a(F_\theta)\chi_{M_\theta\leq \rho}\|&\leq&
\frac{1}{\sqrt{\sin\delta'}} \,  \rho^{1/2} \, ||F||_\rho,
\label{relbnd}\\
\left|\scalprod{\psi}{a^{\#}(F_\theta)\psi}\right|&\leq&
\frac{1}{\sqrt{\sin\delta'}} ||F||_\infty ||\psi||\,
||M_\theta^{1/2}\psi||, \label{relbnd'}
\end{eqnarray}
for all $\psi\in{\cal D}(M_\theta^{1/2})$, and where $a^\#$
denotes either $a$ or $a^*$. In particular, \fer{relbnd''} --
\fer{relbnd'} (together with \fer{aaa3} below) imply Lemma~\ref{lemma7_3}.
\end{proposition}

{\it Proof.\ } Note that \fer{A3.3} follows from Eqn
\fer{relbnd''} and the relation 
\begin{equation}
\|a^*(G)\psi\|^2 \leq \|G\|^2\
\|\psi\|^2+\|a(G)\psi\|^2.
\end{equation}   
We prove only \fer{relbnd}. Bound
\fer{relbnd''} is obtained in a similar way (see \cite{BFS:QED},
Lemma~I.6) and bound \fer{relbnd'} follows from \fer{relbnd''}.
Set for short $P_\rho=\chi_{M_\theta\leq\rho}$. We have for any
$\psi$
\begin{equation}
\|a(F_\theta)P_\rho\psi\|^2\leq\left[ \sum_\alpha\int_X
  \|F_\theta(x,\alpha)\|\ \|a(x,\alpha)P_\rho\psi\|\right]^2.
\label{m20}
\end{equation}
Using the pull-through formula
\begin{equation*}
a(x,\alpha) M_\theta = (M_\theta+\sin\delta' |u| +\tau')
a(x,\alpha), 
\end{equation*}
where $x=(u,\sigma)$, we obtain
\begin{equation*}
a(x,\alpha) P_\rho =\chi_{M_\theta+\sin\delta' |u| +\tau'\leq
\rho}\ a(x,\alpha).
\end{equation*}
Because $M_\theta\geq 0$, the integration in \fer{m20} is
restricted to the domain
\begin{equation*}
X_\rho:= \{u\in \BR\ |\ \sin\delta' |u|+\tau'\leq\rho\}\times S^2.
\end{equation*}
Using H\"older's inequality, we obtain from \fer{m20}
\begin{equation*}
\|a(F_\theta)P_\rho\psi\|^2\leq
\left(\sum_\alpha\int_{X_\rho}\frac{\|F_\theta(x,\alpha)\|^2}{|j_\theta(u)|}\right)
\scalprod{P_\rho\psi}{\sum_\alpha\int_{X_\rho}
a^*(x,\alpha)|j_\theta(u)|
  a(x,\alpha) P_\rho\psi}.
\end{equation*}
Since $|j_\theta(u)|\leq |u|+\tau'$, it is clear that the scalar
product on the right side is bounded from above by
\begin{equation*}
\scalprod{P_\rho\psi}{(\Lambda+\tau'N)P_\rho\psi}\leq
 \frac{\rho}{\sin\delta'}   \|P_\rho\psi\|^2.
\end{equation*}
Then, (\ref{relbnd}) follows from definition (\ref{ir}). \hfill\qed

Observe that we have, for any $\nu>1/2$,
\begin{equation} \label{eqn_A.3.6}
\norm{F}_\rho \le \left( \frac{2\rho}{\sin\delta'} \right)^{\nu
-1/2} ||| F |||_\nu,
\end{equation}
and
\begin{equation}
\|F\|_\infty=|||F|||_{1/2},
\label{eqn_A.3.6'}
\end{equation}
where we define 
\begin{equation} \label{eqn_A.3.7}
||| F |||_\nu := \left( \sum_\alpha \int\limits_{\BR\times S^2}
\frac{\norm{F_\theta(x,\alpha)}^2}{|j_\theta(u)|^{2\nu}} du d\sigma
\right)^{1/2}.
\end{equation}
A bound on the norms $|||F_{1,2}|||_\nu^2$, where $F_{1,2}$ are given in \fer{eqn_def_F1}, \fer{eqn_def_F2}, in terms of $\|G_{1,2}\|_{\mu,\theta}$, \fer{eqn_cond_B2}, is obtained as follows. First one sees that for $z=j_\theta(u)=e^{\delta{\rm sgn}(u)}u+\tau$, $|\rIm\delta|<\delta_0$, $|\tau|<\tau_0$, $\tau_0/\cos\delta_0<2\pi/\betamax$ (where $\betamax=\max(\beta_1,\beta_2)$), one has
\begin{equation}
\frac{|z|}{|e^{\beta' z}-1|} \leq 2|z| +\frac{C}{\beta'},
\label{aaa1}
\end{equation}
for all $\beta'\leq\betamax$, and 
where $C$ is a constant which depends only on $\tan\delta_0$. Using this bound in \fer{eqn_def_F1} gives 
\begin{eqnarray}
\lefteqn{
\|F_1(j_\theta(u),\sigma,\alpha)\|^2}\label{aaa2}\\
&& \leq C(1+1/\beta_\alpha) \max_{k=1,\ldots,4}\left\|\gamma\left[\sqrt{|u|+1}\ G_{\alpha k}\right](j_\theta(u),\sigma)\right\|^2,
\nonumber
\end{eqnarray}
where we recall that $\gamma$ was defined in \fer{gamma2}. Estimate \fer{aaa2} implies
\begin{eqnarray}
|||F_1|||_\nu^2&\leq &C\sum_{j=1,2}\sum_{k=1,3} (1+1/\beta_j)\int_{{\mathbb R}\times S^2}dud\sigma \left\|\gamma_\theta\left[ \frac{\sqrt{|u|+1}}{|u|^\nu}G_{jk}\right](u,\sigma)\right\|^2\nonumber\\
&\leq& C\sum_{j=1,2}(1+1/\beta_j)\|G_j\|^2_{\nu,\theta},
\label{aaa3}
\end{eqnarray}
where $\|G_j\|_{\nu,\theta}$ is given in \fer{eqn_cond_B2}. 
The same bound is obtained for $|||F_2|||^2_\nu$.


\section{Level Shift Operator}
\label{appC'}

We prove estimate \fer{eqn_7.31}.  We pass to
the Jak\u si\'c-Pillet glued Hilbert space representation (see Appendices \ref{AppC} and \ref{b2}) and omit
the tilde over the operators.  In the definition
\begin{equation}
\Lambda_{e\rho\theta}:=P_{e\rho} I_\theta \overline P_{e\rho}
L^{-1}_{0\theta}\overline P_{e\rho}I_\theta P_{e\rho}
\end{equation}
we substitute expression \fer{itheta} for the operator $I_\theta$ and,
using the pull-through formulae, pull the annihilation operators
to the right and the creation operators to the left until they
stand next to the operators $P_{e\rho}$.  As a result we obtain
the decomposition
\begin{equation}
\Lambda_{e\rho\theta}=\Lambda^{\rm contracted}_{e\rho\theta} + R\
,
\end{equation}
where $\Lambda^{\rm contracted}_{e\rho\theta}:=P_{e\rho} \av{I_\theta
\overline P_{e\rho}L^{-1}_{0\theta} I_\theta} P_{e\rho}$ is the
contracted term and the term $R$ consists of remaining terms. Here, we use the notation
\begin{equation*}
\av{I_\theta f(\Lambda,L_r)I_\theta} = \av{I_\theta f(\Lambda+\lambda,L_r+\ell)I_\theta}_\Omega|_{\lambda=\Lambda, \ell=L_r},
\end{equation*}
where $\av{\cdot}_\Omega={\rm Tr}_{{\cal F}}(\cdot P_\Omega)$, $P_\Omega$ is the projection onto ${\mathbb C}\Omega$ (the vacuum sector in ${\cal F}$), and where $f$ is a function of two variables.

The remaining terms, $R$, are estimated using \fer{relbnd} and \fer{m20} and
$\|P_{e\rho} L^{-1}_{0\theta} P_{e\rho}\|\leq c\rho^{-1}$.  For
instance one of the terms appearing in $R$ is of the form
\begin{equation}
P_{e\rho} a^*(F_{i\theta}) \overline P_{e\rho} L^{-1}_{0\theta}
\overline P_{e\rho} a(F_{j\theta}) P_{e\rho}
\end{equation}
which is bounded by (see \fer{relbnd} and \fer{m20})
\begin{eqnarray*}
\lefteqn{
\| P_{e\rho} a^*(F_{i\theta})\| \ \|\overline P_{e\rho}
L^{-1}_{0\theta} \overline P_{e\rho}\| \ \|a(F_{j\theta})
P_{e\rho}\|}\\
&\leq & \left(\frac{c}{\rho\sin\delta'}\right)^{1/2} \|F_i\|_p \
c\rho^{-1}
\left(\frac{c}{\rho\sin\delta'}\right)^{1/2} \|F_j\|_\rho \\
&\leq & \left(\frac{c}{\rho\sin\delta'}\right)^{1/2}
\left(\frac{c}{\sin\delta'}\right)^{\mu-1/2} \|F_i\|_\mu
c\rho^{-1}
\left(\frac{c}{\rho\sin\delta'}\right)^{1/2}\left(\frac{c}{\sin\delta'}\right)^{\mu-1/2}
\|F_j\|_\mu.
\end{eqnarray*}
Similarly, we estimate other terms in $R$ to obtain
$R=O(g^{2\mu-1})$.  Now, using $\overline{P}_{e\rho}=\id-P_{e\rho}$ we write
the operator $\Lambda^{\rm contracted}_{e\rho\theta}$ as
\begin{equation}
\label{lcontracted} \Lambda^{\rm contracted}_{e\rho\theta} =
\Lambda'_{e\rho\theta}+\Lambda''_{e\rho\theta}
\end{equation}
where $\Lambda'_{e\rho\theta}:=P_{e\rho} \av{I_\theta L^{-1}_{0\theta} I_{\theta}}
P_{e\rho}$ and
\begin{equation}
\Lambda''_{e\rho\theta}= -P_{e\rho} \av{I_\theta P_{e\rho} L^{-1}_{0\theta}
I_\theta} P_{e\rho}\ .
\end{equation}
Note that both terms on the r.h.s. of (\ref{lcontracted}) are
well-defined since $I_\theta(\psi\otimes\Omega)$ is orthogonal to
${\rm Null} (L_{0\theta})$, for all $\psi\in{\cal H}_p\otimes{\cal H}_p$. A simple computation shows that
$\Lambda''_{e\rho\theta}$ is equal to $P_{e\rho}$ times an integral over
$\omega\leq\rho$ of the trace of the product of two coupling
functions $F_{j\theta}$ divided by a function of the form
$\pm\cosh\delta\omega + \sinh\delta\omega + \tau$ which is bounded
below by $c\sin\delta'\omega$.  Hence that integral is
bounded by $c\rho^{2\mu-1}
\left(\sum_j\|G_j\|_{\mu,\theta}\right)^2$ and,
consequently, $\Lambda''_{e\rho\theta}=O(\rho^{2\mu-1})$ 

A simple consideration shows that $\av{I_\theta L^{-1}_{0\theta}I_\theta}$ is independent of $\theta$, and $\Lambda'_{e\rho\theta}-\Lambda_e P_{e\rho}$ is of order $O(\rho^{2\mu-1})$ as well. Hence, 
\begin{equation}
\Lambda_{e\rho\theta}=\Lambda_eP_{e\rho}+O(\rho^{2\mu-1})\ .
\end{equation}

\end{document}